\documentclass[twocolumn,trackchanges,twocolappendix]{aastex631}

\usepackage{amsmath}
\usepackage{makecell}
\usepackage{graphicx}
\usepackage{amsthm,amsmath,amssymb} 
\usepackage{mathrsfs}
\usepackage{longtable}
\usepackage{subfigure}
\usepackage{rotating}
\usepackage{amssymb}
\usepackage{epsfig}
\usepackage{multirow}
\usepackage{enumerate}
\usepackage{float}
\usepackage[section]{placeins}
\usepackage{url}
\usepackage{soul}
\usepackage{CJK}
\usepackage{booktabs}
\usepackage{comment}

\def\Tot{Total simulated injections}
\def\Inj{Successful injections}
\def\Det{Detected injections}
\def\Rec{Reconstructions}

\def\tot{total simulated injections}
\def\inj{successful injections}
\def\det{detected injections}
\def\mea{Catalog 1 measurements}
\def\rec{reconstructions}

\begin{document}
\begin{CJK*}{UTF8}{gbsn}

\title{Bias-corrected Fast Radio Bursts Population and Spectra Using CHIME Injection Data}

\author[0000-0002-6165-0977]{Xiang-han Cui}
\affiliation{National Astronomical Observatories, Chinese Academy of Sciences, Beijing 100101, China}
\affiliation{International Centre for Radio Astronomy Research, Curtin Institute of Radio Astronomy, Perth 6102, Australia}
\affiliation{School of Astronomy and Space Science, University of Chinese Academy of Sciences, Beijing 100049, China}
\email{cuixianghan@nao.cas.cn}

\author[0000-0002-6437-6176]{Clancy. W. James}
\affiliation{International Centre for Radio Astronomy Research, Curtin Institute of Radio Astronomy, Perth 6102, Australia}
\email{clancy.james@curtin.edu.au}

\author[0000-0003-3010-7661]{Di Li}
\affiliation{Department of Astronomy, Tsinghua University, Beijing 100084, China}
\affiliation{National Astronomical Observatories, Chinese Academy of Sciences, Beijing 100101, China}

\author[0000-0003-1908-2520]{Cheng-min Zhang}
\affiliation{National Astronomical Observatories, Chinese Academy of Sciences, Beijing 100101, China}
\affiliation{School of Astronomy and Space Science, University of Chinese Academy of Sciences, Beijing 100049, China}
\affiliation{School of Physical Sciences, University of Chinese Academy of Sciences, Beijing 100049, China}

\begin{abstract}
Fast Radio Bursts (FRBs), a class of millisecond-scale, highly energetic phenomena with unknown progenitors and radiation mechanisms, require proper statistical analysis as a key method for uncovering their mysteries. In this research, we build upon the bias correction method using pulse injections for the first CHIME/FRB catalog, to include correlations between properties, and to analyze the FRB population spectrum. This model includes six FRB properties: dispersion measure (DM), pulse width, scattering timescale, spectral index, spectral running, and fluence. By applying the multidimensional weight function calculated by the model, we update the corrected distributions, suggesting that more low-DM, short and long-width, and short-scattering timescale events may exist.
Using one-off events and the first bursts from repeaters, the derived intrinsic population spectrum has a best-fit power-law of $F(\nu)\propto\nu^{\alpha}$, where $\alpha=-2.29\pm0.29$. This confirms previous indications that FRBs are brighter or more numerous at low frequencies. Analysing non-repeaters only, we find $\alpha=-2.50\pm0.43$, while including all bursts from repeaters produces $\alpha=-1.91\pm0.20$. This hints that active repeaters, low-rate repeaters, and non-repeaters may have different progenitors, mechanisms, or evolutionary stage.

\end{abstract}


\section{Introduction} \label{1}
Fast Radio Bursts (FRBs), highly energetic radio bursts lasting only a few milliseconds and originating from cosmological distances, have captured significant attention from the astronomical community since their discovery \citep{Lorimer07}. 
After years of rapid development, numerous milestones have been made \citep{Cordes19, Petroff22, Zhang23}. 
To date, approximately 800 FRB sources and nearly 10,000 bursts have been released\footnote{https://blinkverse.zero2x.org}. 
The Canadian Hydrogen Intensity Mapping Experiment (CHIME) \citep{CHIME18}, with its large field of view, has discovered the majority of the known FRB sources \citep{CHIME21a}. 
This provides an excellent data sample and a unique opportunity for statistical and population analysis of FRBs, and the upcoming new catalog from CHIME will include over 3,000 individual sources\footnote{Information from the Fast Radio Burst 2024, Seth Siegel's talk}.

Frequency spectral properties are a key component of FRB studies \citep{Macquart19}, as they illustrate how fluence varies with frequency and event rate, providing important clues to clarify their origins and mechanisms \citep{Law17}.
The frequency dependence of FRBs helps to identify and probe the surrounding environments and propagation effects along the path, such as scattering or absorption features \citep{Masui15, Prochaska19, Ocker21}.
Additionally, FRB spectral behaviour is critical to understanding their evolution with redshift \citep{James22}.
Therefore, spectral properties are crucial for refining their population characteristics, constraining their progenitors and environments, and enhancing the use of FRBs as cosmological tools.

As the amount of data increases, many studies have focused on analyzing the frequency spectral properties of FRBs.
Studies on active FRB sources have shown that they exhibit a variety of indices \citep{Law17, Houben19, Chawla20, Zhang23b, Ikebe23}.
Various factors may affect the complexity of FRB spectral characteristics and their frequency dependence. 
For example, calibration differences between telescopes with different beam widths \citep{CHIME19c}, absorption and radio frequency interference (RFI) \citep{Rajwade20b}, dispersion and scattering smearing at low frequencies, human selection biases \citep{2018MNRAS.474.1900M}, as well as the intrinsic diversity of FRB sources \citep{Fonseca20, Cui21}.
Therefore, using data from the same telescope with consistent calibration to analyze the statistical spectra of FRB population becomes essential.

However, research in this area remains relatively limited.
Two methods currently constrain spectral behaviour: measurements of the frequency dependent fluence of FRBs, and FRB population models.
Assuming all FRBs are broad-band events and follow a power-law population spectrum of $F(\nu)\propto \nu^{\alpha}$, \cite{Macquart19} fit the frequency dependent fluence of 23 FRBs detected by the Australian SKA Pathfinder (ASKAP) to obtain a mean spectral index of $\alpha=-1.5^{+0.2}_{-0.3}$ (\citet{James22} notes that if all FRBs are extremely narrow-band, the result of \cite{Macquart19} should be corrected to $\alpha=-0.65\pm0.3$).
Fits to the redshift--DM distribution of FRBs can also constrain the spectral index, though as both \cite{James22} and \cite{Bhattacharyya22} note, such population modeling of spectral evolution is largely degenerate with source evolution. 
Using data from ASKAP and Murriyang (Parkes), 
\cite{James22b} found a best-fitting value of $\alpha=-1.0\pm0.85$, while using CHIME data, \cite{Shin23} derived an index of $\alpha = -1.39^{+0.86}_{-1.19}$, assuming all events share similar spectral properties.
Thus, it is imperative to determine the spectral behaviour of FRBs to improve the accuracy of FRB population and cosmological studies.
It should be noted that the $\alpha$ in this work refers to the index of the frequency-fluence relation, not the index of the energy distribution.

The first CHIME/FRB catalog plays an important role since its release \citep{CHIME20a}, and statistical analyses of the CHIME/FRB data have been widely conducted \citep{Chawla22, Pleunis21b, Hashimoto22, Cui22, Shin23, Wang24}. 
Nevertheless, research on population spectra using CHIME/FRB Catalog 1 data is still absent.
Potentially, this is because of
significant uncertainties of spectral parameters and fluence measurements \citep{Andersen23}, selection effects, and complex instrument response and beamshape \citep{Amiri22a}. 
\cite{Merryfield23} developed an injection system with bursts characterized by dispersion measure (DM), spectral index, pulse arrival time, and pulse width. This system was utilized by CHIME/FRB Catalog 1, which used the injection results to correct the measured distributions of DM, scattering timescale, and pulse width. 
Additionally, \cite{Shin23} applied it to derive the energy and distance distribution.

In this work, we further expand the correction approach of CHIME/FRBs \citep{Merryfield23, Shin23} by using injection data to construct a statistical model that additionally considers parameter correlations, obtaining updated bias-corrected population.
We also develop a correction for the frequency spectrum of CHIME/FRBs and derive the intrinsic population spectrum.
The paper is organized as follows.
In Section \ref{2}, we provide a detailed introduction to the measurements from the first CHIME/FRB catalog and the injection data, explaining our approaches.
In Section \ref{3}, the construction steps of the statistical model are explained in detail, along with the method for calculating the multi-dimensional weight function.
In Section \ref{4}, we apply the weight function derived from the model to correct the FRB distributions and population spectra.
In Section \ref{5}, we discuss the reliability of the results and their implications. 
Finally, a brief conclusion is presented in Section \ref{6}.



\section{Data sets and approach} \label{2}

Several data sets have been used for the following model construction and calculations, which are from the first CHIME/FRB catalog \citep{CHIME21a} (hereafter refer to as Catalog 1) and its injection system \citep{Merryfield23}. 
Since the analysis involves dealing with selection effects, correlations between parameters, and beam shape, the details of the data preparation need to be clarified in the beginning.

\subsection{The first CHIME/FRB catalog and its injections}\label{2.1}
The CHIME/FRB Collaboration has released Catalog 1 with 536 FRBs in 2021, including 474 one-off events (apparent non-repeaters, hereafter as non-repeaters) and 62 bursts from 18 repeating FRBs \citep{CHIME21a}\footnote{http://www.chime-frb.ca}.
Unlike other studies, to keep more data and avoid introducing additional human biases, we used the full data set in our main text.
Since the non-repeaters and the first burst of repeaters in CHIME have the same detection algorithm, we used the 474 non-repeaters and the first burst of 18 repeaters as our standard sample of \mea, as per \citet{Shin23}.
Meanwhile, to ensure the completeness of our analysis, we also considered the case with data cuts \citep{CHIME21a, Shin23} in Appendix \ref{A}, demonstrating that whether or not cuts are applied to data does not significantly affect our conclusions.
For each burst, six properties are used in our population studies \citep{CHIME21a}: dispersion measure (DM), burst width ($w$), scattering timescale ($\tau$), spectral index ($\gamma$), spectral running ($r$), and fluence ($F$).
In the Catalog 1 database, these FRB properties correspond to columns of \texttt{dm\_fitb}, \texttt{width\_fitb}, \texttt{scat\_time}, \texttt{sp\_idx}, \texttt{sp\_run}, and \texttt{fluence}, respectively.
The values of these data are determined  by the fitting algorithm named \texttt{fitburst}, and the scattering timescale is transformed to 600 MHz by assuming $\tau \propto \nu^{-4}$.
Comparing to other data sets, CHIME/FRB Collaboration use two parameters ($\gamma$ and $r$) to describe FRB spectral properties \citep{Pleunis21b} with fluence
\begin{equation}
    F(\nu)\propto (\nu/\nu_0)^{\gamma+rln(\nu/\nu_0)},
    \label{fgr}
\end{equation}
where $\nu_0$ is defined as a fixed constant of 400.1953125 MHz.
The reason for using two parameters instead of the usual one to describe the spectra is that many FRBs are frequency narrow-band. 
So, adding the running parameter ($r$) allows for a better description of the limited frequency band.

Considering possible biases of instrument selection effects, RFI removal algorithms, and the detection pipeline, CHIME developed a real-time injection system \citep{Merryfield23}\footnote{https://chime-frb-open-data.github.io/injections/}.
Through injecting synthetic FRBs generated by the simulation code \texttt{simpulse}\footnote{https://github.com/kmsmith137/simpulse} into the pipeline, selection effects can be characterised by comparing the ratio of detected to injected bursts as a function of burst properties.
Hereafter, we refer to these synthetic injection data as injections.
Specifically, a total number of $5\times10^6$ simulated FRBs were generated independently, among them $\sim 9.7\times 10^4$ (96942) bursts passed the S/N estimation and prepared for injection.
Due to network issues, $\sim 8.5\times 10^4$ (84697) bursts were successfully injected, and $\sim 3.0\times 10^4$ (29722) were detected by the system with thresholds of SNR $>$ 9 and RFI grade $>$ 7, or an SNR $>$ 30.
Here, both filtering made prior to \inj, and the process determining which are \det, can be considered as selection effects \footnote{Note that the CHIME injection catalog does not directly provide six properties for each \det. 
However, detected events can be matched to their injected properties by comparing the detection IDs.}.
It is worth noting that the parameter spaces of \tot\ are much broader than those of \mea.
For instance, the scattering timescale range in \tot\ ($2.6\times10^{-8}\,s - 37\,s$) is significantly wider compared to \mea\ ($1.0\times10^{-4}\,s - 9.0\times10^{-2}\,s$). 
Therefore, we believe that \tot\ can approximately represent the full parameter space of all FRBs within the CHIME frequency band ($\mathrm{400~MHz-800~MHz}$).
A brief summary of the data sets and their abbreviations that will be used are shown in the following points:
\begin{itemize}
    \item \Tot\ (tot): a total of $5\times10^6$ simulated injections are represented as $(\mathrm{DM},\,w,\,\tau)_{\mathrm{tot}}$, $(\gamma,\,r)_{\mathrm{tot}}$, and $F_{\mathrm{tot}}$.
    
    \item \Inj\ (inj): $\sim 8.5\times 10^4$ successfully injected injections are illustrated as $(\mathrm{DM},\,w,\,\tau)_{\mathrm{inj}}$, $(\gamma,\,r)_{\mathrm{inj}}$, and $F_{\mathrm{inj}}$.
    
    \item \Det\ (det): $\sim 3.0\times 10^4$ detected injections after the pipeline are described as $(\mathrm{DM},\,w,\,\tau)_{\mathrm{det}}$, $(\gamma,\,r)_{\mathrm{det}}$, and $F_{\mathrm{det}}$. 
    
    \item \mea\ (mea): the actual measured values of $(\mathrm{DM},\,w,\,\tau)_{\mathrm{mea}}$ presented in Catalog 1, including 474 non-repeaters and the first burst of 18 repeaters.
    
    \item \Rec\ (rec): reconstructed data of $(\gamma,\,r)_{\mathrm{rec}}$ and $F_{\mathrm{rec}}$ (see Section \ref{2.3}).
\end{itemize}

\begin{figure*}
\plotone{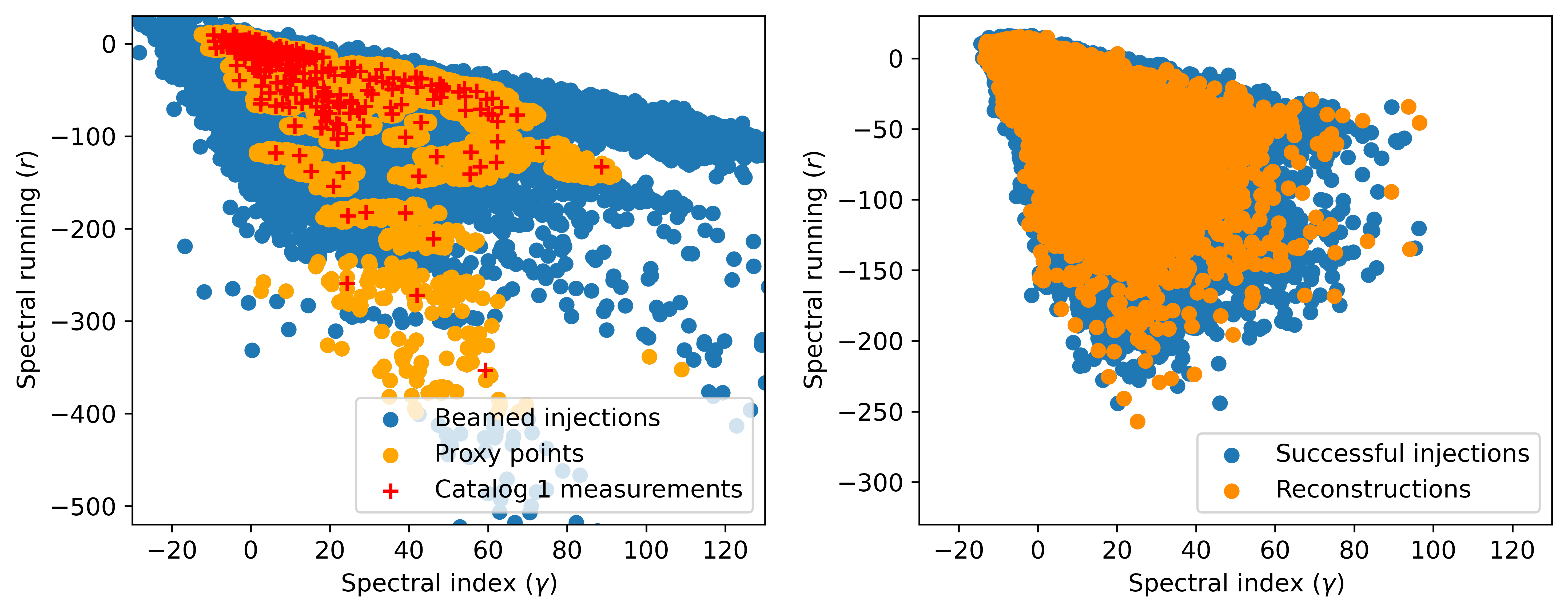}
\caption{ 
Distributions of spectral index ($\gamma$) and spectral running ($r$) for reconstructions.
Left panel: illustration of reconstruction method. Blue points represent detected injections modulated by the beam. 
Orange points mean proxy points of reconstructed data, which are selected from the 50 nearby \mea.
Red pluses indicate the \mea. 
Right panel: distribution of \rec\ within \inj. Blue points show the \inj. 
Orange points are \rec, mapping from the proxy points in the left panel.}
\label{rec-gr}
\end{figure*}

\subsection{Approach} \label{2.2}
The correlations of both intrinsic FRB parameters and observational selection effects are important issues in the process of statistical modeling.
In previous studies, correlations were often considered as negligible, or having a minimal impact on the overall distribution \citep{CHIME21a}.
In our approach, we assume that correlations exist inside the $(\mathrm{DM},\,w,\,\tau)$ and $(\gamma,\,r)$ sets, but not between $(\mathrm{DM},\,w,\,\tau)$, $(\gamma,\,r)$, and $F$.
Thus, the likelihood of an FRB with six properties is described as 
\begin{equation}
    P(\mathrm{DM}, w, \tau, \gamma, r, F) = P(\mathrm{DM}, w, \tau) P(\gamma, r)P(F).
    \label{pmulti}
\end{equation}
Here, we do not use the full 6-dimensional analysis because the dataset is not large enough, so separating frequency-dependent effects from time-dependent effects is a prudent assumption, which means we only consider the correlations within (DM, $w$, $\tau$) and ($\gamma$, $r$), rather than between them.
Meanwhile, since we will apply weights to the \tot, this means that the weighted distribution we obtain will also not exhibit any correlations between $(\mathrm{DM},\,w,\,\tau)$, $(\gamma,\,r)$, and $F$. 
Therefore, the correlation of selection effects should also be chosen to reflect any potential correlation between the parameters of the underlying FRB population.

Obviously, the correlation between $\gamma$ and $r$ is understandable, because they are determined by Eq. \ref{fgr}.
For FRB properties of DM, $w$, and $\tau$, intrinsic correlations may arise from propagation through the medium along the line of sight \citep{Cordes16b, Qiu20}, in particular, DM and scattering \citep{Ocker21}. Observationally, dispersion and scattering both lead to pulse broadening as
\begin{equation}
    w\propto\sqrt{w_o^2+w_{\mathrm{DM}}^2+w_{\tau}^2},
    \label{wsim}
\end{equation}
where $w_o$ means the intrinsic width, and $w_{\mathrm{DM}}$ and $w_{\tau}$ represents pulse broadening caused by dispersion and scattering, respectively.
Thus, the broadening caused by these three parameters may be degenerate with each other.
To illustrate this, we show Pearson correlation coefficients and P-values \citep{Pearson95} for the DM, $w$, and $\tau$ from \mea\ in Table \ref{table1}.
In the Pearson test, the null hypothesis states that there is no correlation. 
Correlation coefficients significantly different from zero are evidence against this hypothesis, with the quoted p-values bring the probability that the correlation coefficient is at least that large assuming the null hypothesis to be true.
All three cases in Table \ref{table1} show positive correlations, though the correlation between DM and $\tau$ is only marginally significant.
The observed positive correlations could be due to selection effects, intrinsic differences, or a combination of both.
\begin{table}[h!]
\centering
\caption{Results of statistical test on correlation of parameters within \mea}
\begin{tabular}{ccc}
\toprule
\toprule
Properties$^a$  & Correlation coefficient &  P-value$^b$ \\
\midrule
DM and $w$  & $1.2\times10^{-1}$ & $8.1\times10^{-3}$ \\
DM and $\tau$  & $8.7\times10^{-2}$  & $5.5 \times10^{-2}$ \\
$w$ and $\tau$  & $4.1\times10^{-1}$  & $1.5\times10^{-21}$ \\
\bottomrule
\end{tabular}
\label{table1}
\begin{flushleft}
Notes. $^a$ The three parameters are dispersion measure (DM), pulse width ($w$), and scattering timescale ($\tau$) from \mea.
$^b$ If the p-value is less than 0.05, the null hypothesis can be rejected at 0.05 significance level.
\end{flushleft}
\end{table}


\subsection{Data reconstruction of  $\gamma$, $r$, and $F$}
\label{2.3}
The accuracy of the data used to construct the model directly impacts the reliability of the results. 
We assume that \mea\ of DM, $w$, and $\tau$ are sufficiently close to the true values.
However, the insufficient localization capability within the beam for Catalog 1 FRBs introduces uncertainty in determining the correct beam response applicable to those FRBs \citep{CHIME21a}.
In turn, this means that the measured values of burst fluence, and fitted spectral properties $\gamma$ and $r$, deviate from their true values. 
Therefore, it is necessary to appropriately reconstruct $\gamma$, $r$, and $F$ in \mea.

To reconstruct $\gamma$ and $r$, we treat the $\gamma$ and $r$ of each FRB event in \mea\ as a two-dimensional array $(\gamma,\,r)_{\mathrm{mea}}$.
Additionally, we know that there are $\sim 9.7\times 10^4$ beam-modulated injected events, and we identify the ones detected by the system $(\gamma,\, r)_{\mathrm{beam}}$.
Then, these $(\gamma,\,r)_{\mathrm{mea}}$ are compared with $(\gamma,\, r)_{\mathrm{beam}}$, as shown in the left panel in Figure \ref{rec-gr}.
The different spectral dependencies of beam sensitivity at different points in the beam mean there is not a 1-1 mapping between measured and true values of $\gamma$ and $r$.
Hence, we select the 50 closest $(\gamma,\,r)_{\mathrm{beam}}$ points in the neighborhood of each $(\gamma ,\, r)_{\mathrm{mea}}$ and use their simulated true values of $\gamma$ and $r$ (i.e. injected values before being modified by the beam shape) to form the reconstructed data as orange points in the right panel of Figure \ref{rec-gr}.

Then, we consider the fluence.
Unlike all other parameters, where we need the data to tell us the truth, here we just assume that true fluence distribution goes as $dNdF^{-1}\propto F^{-2.5}$.
The reason for choosing a distribution index of $-2.5$ is that we assume it follows Euclidean conditions with a flat local universe.
Therefore, the reconstructed $(\gamma,r)$ and $F$ can be written as $(\gamma,\,r)_{\mathrm{rec}}$ and $F_{\mathrm{rec}}$, respectively.

\section{Model construction} \label{3}
\begin{figure}[ht]
\plotone{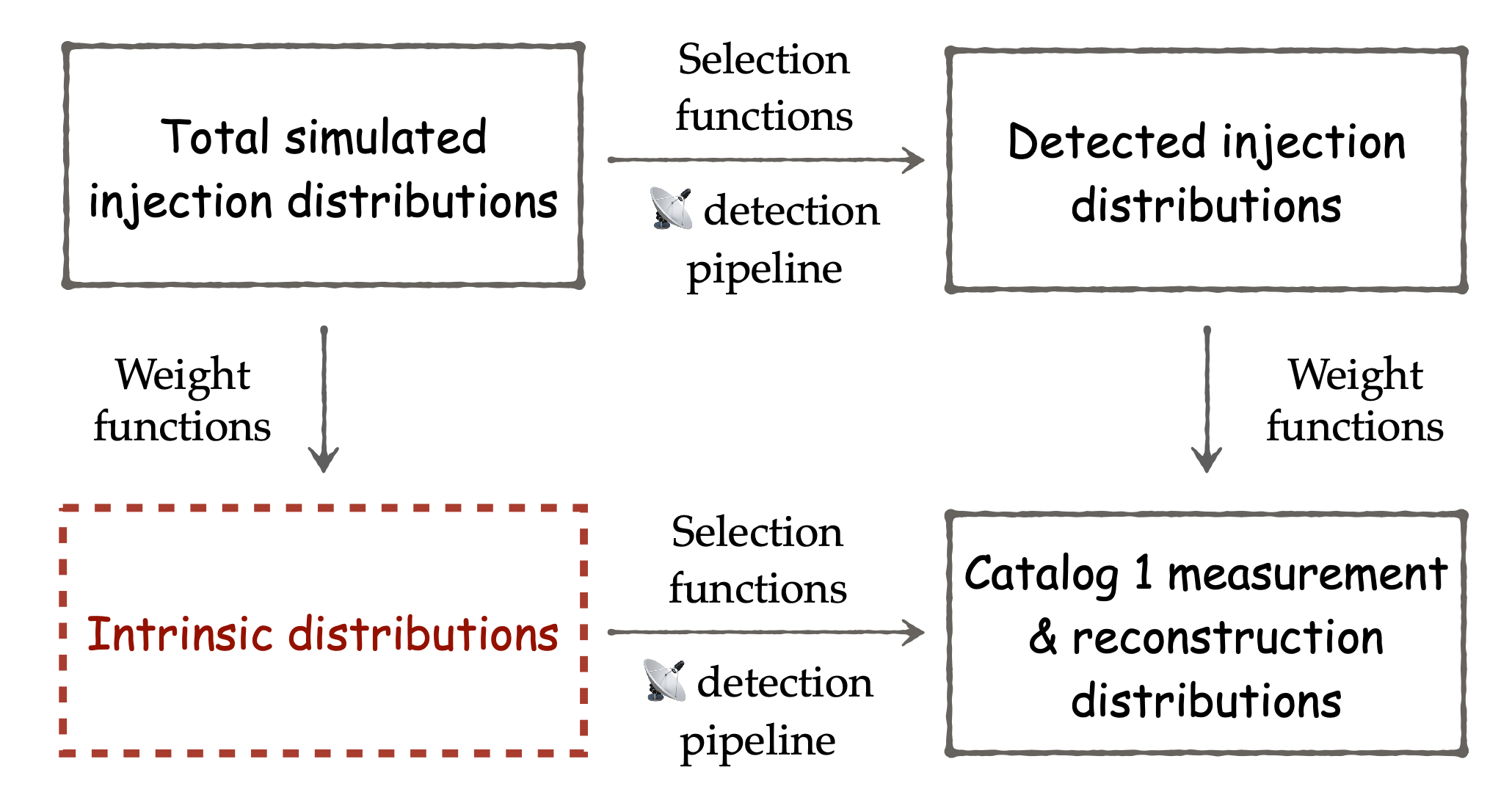}
\caption{Conceptional diagram of the model.
Selection functions describe the selection effects caused by the detection pipeline within datasets like Eqs.~\ref{f5minj}-\ref{fs}.
Weight functions are used to express the transformation between datasets, such as  from the injection datasets to the real cases as Eqs.~\ref{wximea}-\ref{wxiF}.}
\label{method}
\end{figure}

Our methodology is an extension of \cite{Shin23}'s correction method, where they constructed a statistical model using injections and \mea\ to derive selection and weight functions, and hence the properties of the underlying FRB population (for details, see Appendix C in \cite{CHIME20a}). This approach, and the role of these functions, is illustrated in Figure \ref{method}.
Our improvements are in considering the correlations between parameters, thus using a multi-dimensional histogram (Eq.~\ref{pmulti}) to calculate the selection and weight function. 
Additionally, we introduce two spectral parameters and fluence into the analysis.

A selection function ($S(\xi)$) can be used to describe pipeline filtering processes. The pulse injection process involves two selection functions: selection between \tot\ and \inj\ ($\mathrm{tot}\rightarrow \mathrm{inj}$), and selection between \inj\ and \det\ ($\mathrm{inj}\rightarrow \mathrm{det}$), which can be written as
\begin{equation}
    S(\xi)_{\mathrm{tot}\rightarrow \mathrm{inj}} = P(\xi)_{\mathrm{inj}}P(\xi)_{\mathrm{tot}}^{-1},
    \label{f5minj}
\end{equation}
\begin{equation}
    S(\xi)_{\mathrm{inj}\rightarrow \mathrm{det}} = P(\xi)_{\mathrm{det}}P(\xi)_{\mathrm{inj}}^{-1},
    \label{finjdet}
\end{equation}
where the parameter $\xi$ represents different multi-dimensional FRB properties of $(\mathrm{DM},\,w,\,\tau)$, $(\gamma,\,r)$ and $F$.
Combining the above two equations, the overall selection function is given by 
\begin{equation}
    S(\xi) = P(\xi)_{\mathrm{det}}P(\xi)_{\mathrm{tot}}^{-1}.
    \label{fs}
\end{equation}
Therefore, the selection functions can be viewed as the ratio of two normalized distributions for each histogram bin.

The selection function can only be calculated directly if the \tot\ was identical to the true underlying FRB population. 
Therefore, weight functions are used to re-weight \det\ to mimic the CHIME Catalog 1 measurements, i.e.\
\begin{equation}
    P(\xi)_\mathrm{mea} = W(\xi)\times P(\xi)_\mathrm{det}.
    \label{recmea}
\end{equation}
If we assume that the injection system can properly mimic the detection efficiency and selection effects of the real-time search pipeline \citep{Merryfield23}, then the true underlying population can then be be conceptually derived from
\begin{equation}
    P(\xi) = W(\xi)\times P(\xi)_\mathrm{tot}.
    \label{defint}
\end{equation}

To determine selection and weight functions, we use the following iterative procedure. In step $i$, we calculate three weight functions using \det\ estimated from the previous step ($i-1$), and \mea\ and \rec\ as
\begin{equation}
    W(\mathrm{DM},\,w,\,\tau)^{i} = \frac{P(\mathrm{DM},\,w,\,\tau)_{\mathrm{mea}}}{P(\mathrm{DM},\,w,\,\tau)_{\mathrm{det}}^{i-1}},
    \label{wximea}
\end{equation}
\begin{equation}
    W(\gamma,\,r)^i = \frac{P(\gamma,\,r)_{\mathrm{rec}}}{P(\gamma,\,r)_{\mathrm{det}}^{i-1}},
    \label{wxirec}
\end{equation}
\begin{equation}
    W(F)^{i} = \frac{F^{-3/2}}{P(F)_{\mathrm{det}}^{i-1}}.
    \label{wxiF}
\end{equation}
Since we assume that correlations only exist inside $(\mathrm{DM},\,w,\,\tau)$ and $(\gamma,\,r )$, respectively, but do not exit between $(\mathrm{DM},\,w,\,\tau)$, $(\gamma,\,r )$, and $F$, the multi-weight function ($W_m$) can be written as
\begin{eqnarray}
    W_m^i & \equiv & W(\mathrm{DM},\,w,\,\tau, \gamma,\,r,\,F)^i \nonumber \\
      &=& W(\mathrm{DM},\,w,\, \tau)^i \times W(\gamma,\,r) \times W(F)^i.
\label{wm}
\end{eqnarray}
For $W(\mathrm{DM},\,w,\,\tau)$ and $W(F)$, we construct multi-dimensional histograms of the data, and calculate the weights independently bin-by-bin.
For $W(\gamma,\,r)$, we choose the number of times that each injection point was selected during the reconstruction process as its weight value.
If a point was not selected, its weight was set to 0.
After that, we multiply these three weights together to obtain the $W_m$ for each event individually.

With this definition, we update our estimates for the intrinsic population as
\begin{eqnarray}
    P(\xi)^{i} & = & P(\xi_{\rm tot}|W_m^i)
    ,\label{pi1}
\end{eqnarray}
which implies an updated \det\ distribution using the selection function in Eq. \ref{fs}:
\begin{eqnarray}
    P(\xi)^i_{\rm det} & = &  S(\xi) P(\xi)^{i}.
    \label{p-deti}
\end{eqnarray}
With our updated \det, we now begin the next iteration by re-calculating the weights (Eqs.\ \ref{wximea}--\ref{wxiF}).
We iterate until convergence is achieved --- typically, 100 iterations sufficed. For our initial estimates ($i=0$), we assume all weights are unity, equivalent to $P(\xi) = P(\xi)_{\rm tot}$.

To derive the intrinsic population, our weights must be applied to all \tot\  --- yet according to the method used to calculate Eq.~\ref{wxirec} and $W(\gamma,\,r)$, only \det\ can carry a non-zero weight. 
Therefore, as a final step, we calculate a modified weight as the ratio between \rec\ and weighted \det\ by histogramming in $(\gamma,\, r)$ space,
\begin{equation}
    W_\mathrm{mod} = \frac{P(\gamma,r)_{\mathrm{rec}}}{ P\big((\gamma,r)_{\mathrm{det}}|W_{m}^{n}\big)}.
    \label{wmodif}
\end{equation}
The superscript $n$ means the multi-weight after n total iterations.
Thus, the final weight function is obtained as
\begin{equation}
    W_\mathrm{final} = W_{m}^{n} \times W_\mathrm{mod}.
    \label{wfinal}
\end{equation}
Therefore, the bias-corrected distributions with the final weights is 
\begin{equation}
    P(\xi)_{i}^{n} = P({\xi}_{\mathrm{tot}}|W_\mathrm{final}).
    \label{piloop}
\end{equation}
Based on this model, we can comprehensively consider various biases within \mea, including selection effects, correlations between parameters, and beam modulations.

\section{Results} \label{4}
\subsection{Corrected distributions of DM, width, and $\tau$}\label{4.1}
\begin{figure}[ht]
\plotone{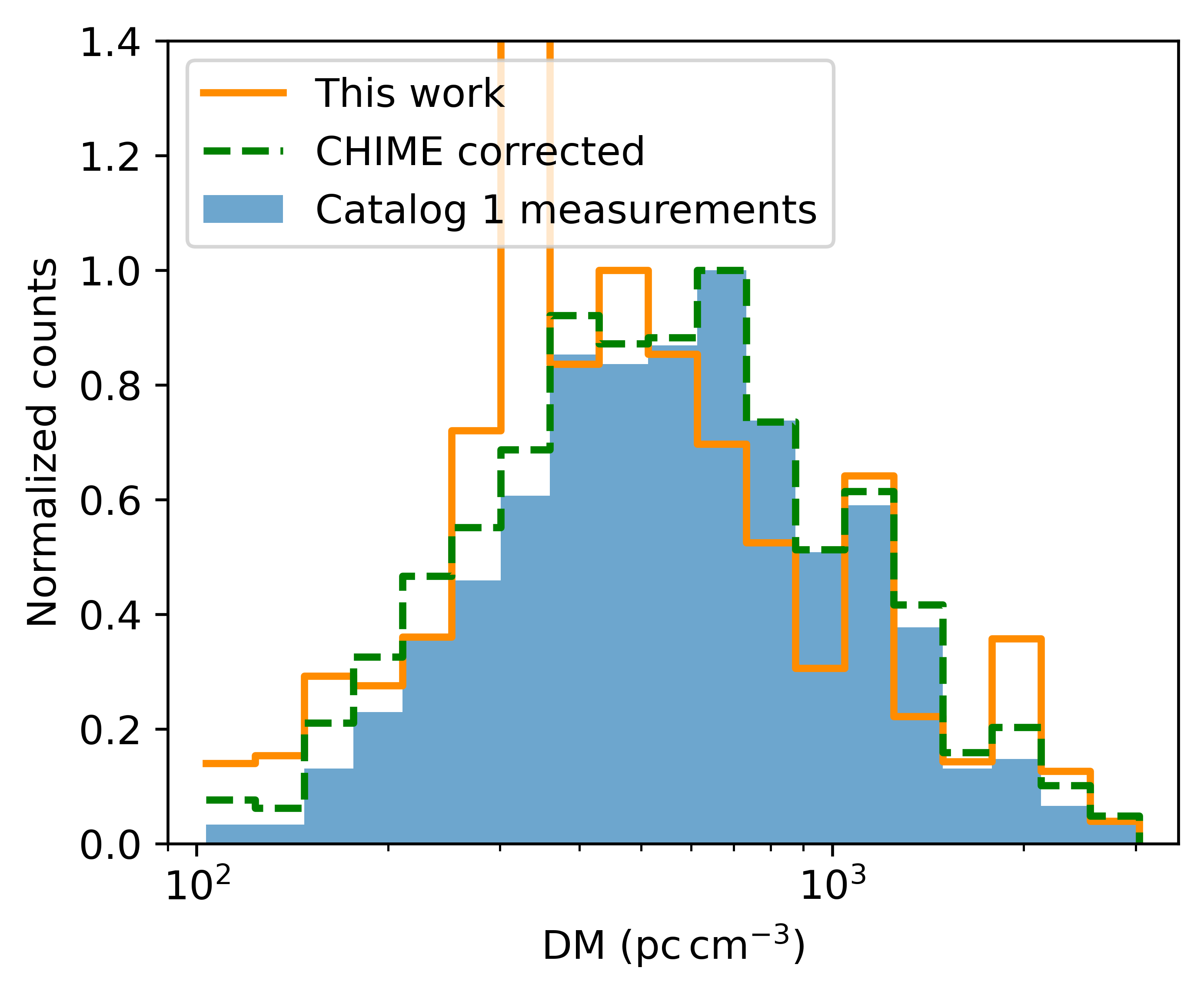}
\caption{Distribution of dispersion measure (DM) with considering different correction method.
The orange solid distribution represents the corrected \tot using the multidimensional weight function in this work.
Here, we do not use the maximum value for normalization, as we consider it might be an outlier. 
Instead, we used the second-highest value for normalization.
The green dashed distribution shows the results with the CHIME selection function.
The blue histogram means the distribution of \mea.
The annotations for the distributions and histograms in Figures \ref{dm-dis}, \ref{w-dis}, and \ref{tau-dis} are the same.}
\label{dm-dis}
\end{figure}

\begin{figure}[ht]
\plotone{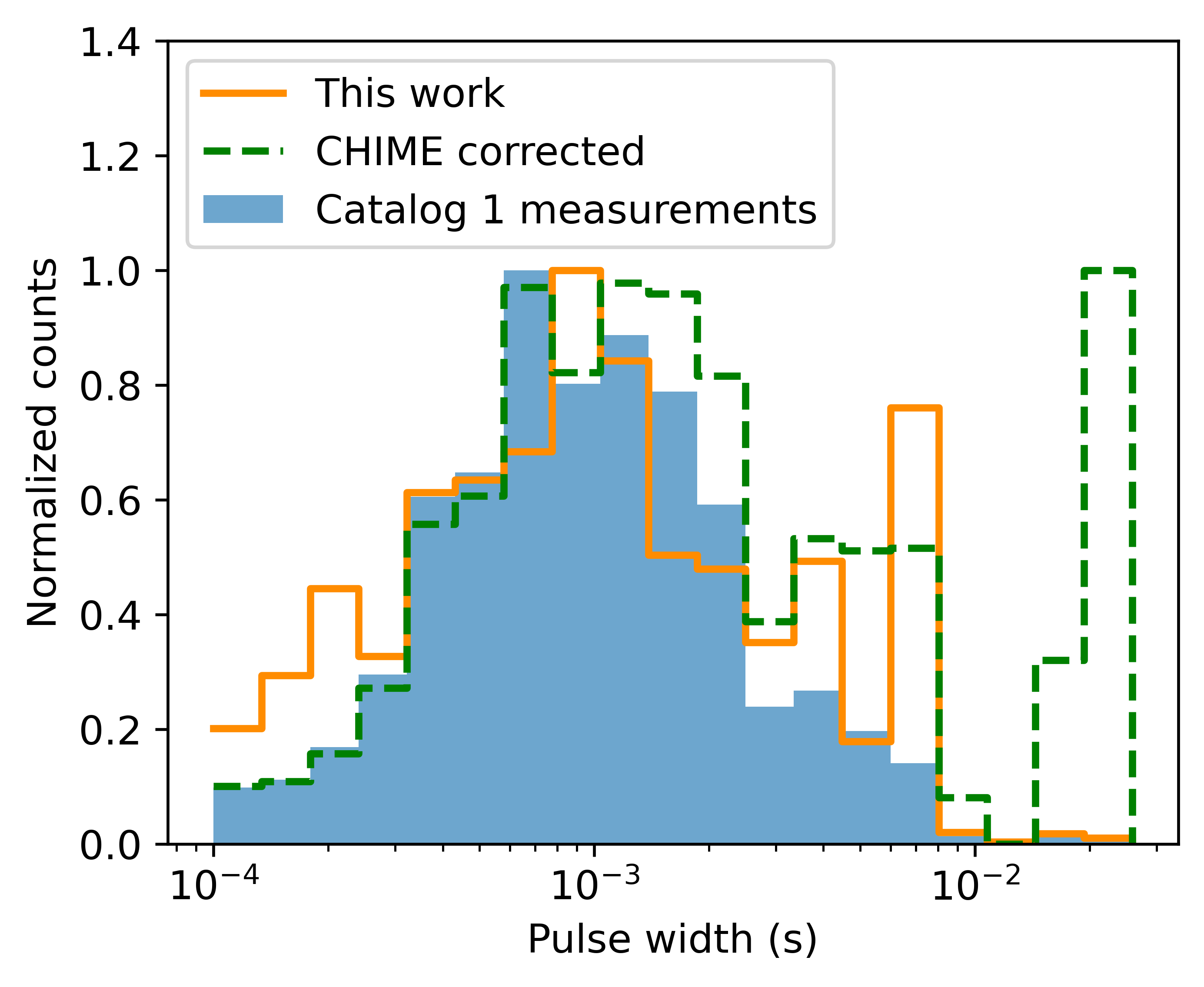}
\caption{Distribution of pulse width ($w$) with considering different correction method. 
Annotations are consistent with those in Figure \ref{dm-dis}.}
\label{w-dis}
\end{figure}

\begin{figure}[ht]
\plotone{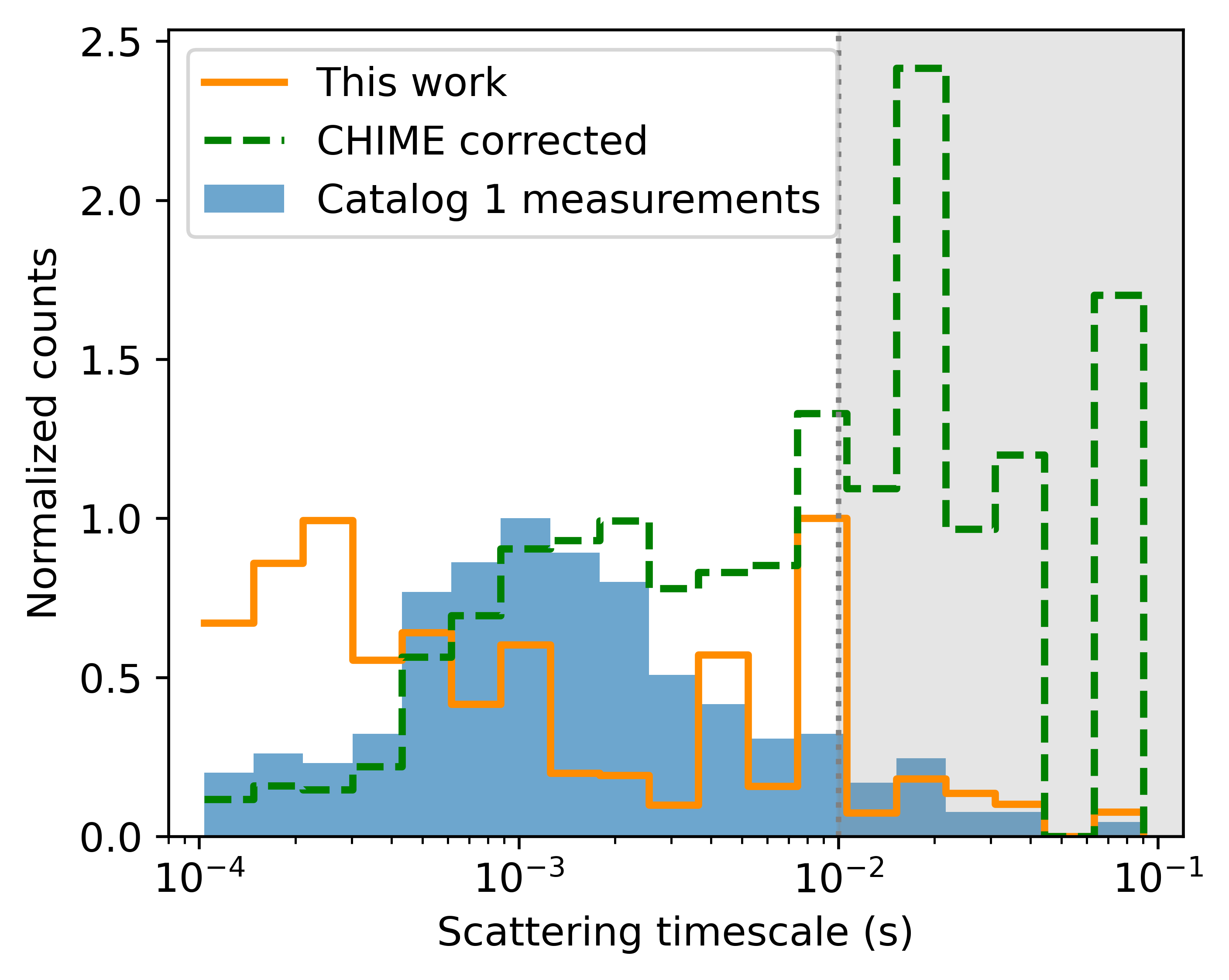}
\caption{Distribution of scattering timescale ($\tau$) with considering different correction method. 
The gray region corresponds to timescales larger than 0.01s, which are poorly constrained in the CHIME correction \citep{CHIME21a}.
Here, we use the maximum value of the bins outside the gray area for normalization.
Other annotations are consistent with those in Figure \ref{dm-dis}.}
\label{tau-dis}
\end{figure}

Using the final multi-weight function Eq. \ref{wfinal}, the distributions of DM, pulse width and scattering timescale can be updated according to Eq. \ref{piloop}.
We present all distributions such that the measured and corrected distributions have the same normalization. 
Thus, regions where the corrected distribution is less than that measured imply relatively less bias against that part of the parameter space.
20-bin histograms are used for each variable in all calculations, and all results are normalized to 1 for comparison.
In Figure \ref{dm-dis}, \ref{w-dis}, and \ref{tau-dis}, we show distributions of \mea, the corrected distributions using our method, and the corrected distribution with the CHIME selection functions for DM, pulse width, and scattering timescale, respectively \citep{CHIME21a}.

In the cases of DM and pulse width in Figure \ref{dm-dis} and \ref{w-dis}, our corrections are broadly the same as the corrections of \citet{CHIME21a}.
The results suggest that there should be more low-DM events, as well as events with shorter and longer duration. 
It is worth noting that an outlier appears in the seventh orange histogram bin from the left in Figure \ref{dm-dis}, which may be due to statistical fluctuations and, hence, has no physical meaning. 
Therefore, we do not use this value for normalization; instead, we use the maximum value excluding this outlier for normalization.
For the scattering timescale case as shown in Figure \ref{tau-dis}, compared to corrections of CHIME, we find significant differences in short and long scattering regions.
CHIME's correction suggests a large number of long-scattering events has been lost by system. However, in our results, the bias in the long-scattering region is relatively small, while there may be more events within short-scattering timescales.

In CHIME's correction, they did not consider ($\gamma,\,r$) in the iteration and assume that other parameters are independent of each other \citep{CHIME21a}.
To test if this is the source of the disagreement, we recalculated the weights as $W(\mathrm{DM}, w, \tau, F) = W(\mathrm{DM}, w, \tau)\times W(F)$, i.e.\ using the same method, but only considering the 4-dimensional case (excluding $\gamma,\,r$). 
The results, which are similar to CHIME's corrections, are discussed further in Appendix \ref{B}. 
We thus attribute the dissimilarities to correlations between ($\gamma,\,r$) and scattering. 
We suggest that a possible source of correlation is because events that are low-frequency biased are significantly more affected by scattering than those which are high-frequency biased, given that scattering is defined as the scattering timescale at 600\,MHz.

\subsection{Corrected spectra}\label{4.2}

\begin{figure*}[ht]
\plotone{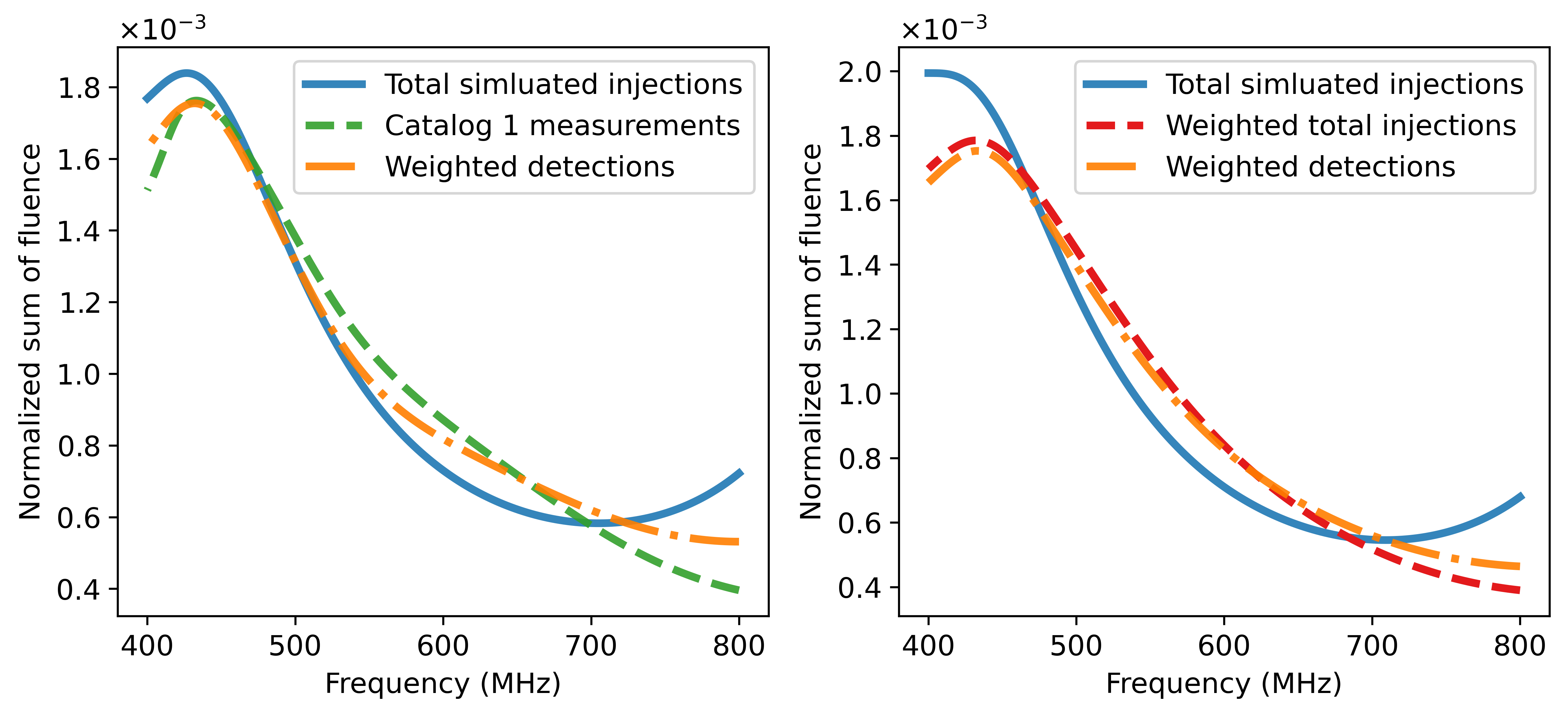}
\caption{The population spectra  with different normalizations.
Left panel:
all the spectra are used Eq. \ref{nor1} for normalization.
The blue solid and  green dashed curves show the spectra of original (unweighted) \tot\ and \mea, respectively.
The orange dash-dotted curve represents the spectrum of weighted \det.
Right panel: all spectra are used Eq. \ref{norflu} for normalization.
The blue solid curve is the spectrum of the original (unweighted) \tot\ with using Eq. \ref{norflu}.
The red dashed curve indicates the spectrum of weighted \tot, while the orange dash-dotted curve represents the weighted spectrum of \det.}
\label{f-v-2}
\end{figure*}



We use final multi-weight function Eq. \ref{wfinal} to re-weight spectra to obtain the corrected spectra.
The cases of unweighted and weighted population spectra can be written as follows,
\begin{equation}
    F(\nu)_{\mathrm{sum}}=\sum_{i=1}^{\mathrm{tot}} F(\nu,\,\gamma_i,\,r_i)_{g}\times \lambda_i,
    \label{unwspectrum}
\end{equation}
and 
\begin{equation}
    F(\nu)_{\mathrm{sum},\,w}=\sum_{i=1}^{\mathrm{tot}} F(\nu,\,\gamma_i,\,r_i)_{g}\times W_{\mathrm{final},\,i}\times \lambda_i,
    \label{wspectrum}
\end{equation}
where $F(\nu,\,\gamma,\,r)_{g}$ is the generated spectrum using $(\gamma,\,r)$ and Eq. \ref{fgr}.
$\lambda$ is a normalization factor, which are calculated using two methods as
\begin{equation}
   \lambda^{\star} =  F_\mathrm{tot}\times \bigg[ \sum_{j=1}^{\mathrm{num}}F(\nu_j,\,\gamma,\,r)_{g}\bigg]^{-1},
    \label{norflu}
\end{equation}
\begin{equation}
   \lambda^{\prime} =  1\times \bigg[ \sum_{j=1}^{\mathrm{num}}F(\nu_j,\,\gamma,\,r)_{g}\bigg]^{-1},
    \label{nor1}
\end{equation}
where superscripts of tot and num is the number of the \tot\ ($5\times10^6$) and frequency channels (1024), respectively.
The first normalization (Eq. \ref{norflu}) ensures that the weight of each FRB is related to its fluence, while the second (Eq. \ref{nor1}) assumes equal weight irrespective of fluence.
We consider both normalization methods
because, as noted in Section \ref{2.3}, the fluence in \mea\ are inaccurate, and fluence distributions of \tot\ and \mea\ are significantly different with $p_{KS}<<0.05$.

We derive the intrinsic FRB spectrum by applying Eq.~\ref{wspectrum} and \ref{norflu} to \tot, which is compared to the unweighted \tot\ and weighted \det\ in the right panel of Figure \ref{f-v-2}. A power-law fit to the weighted total injections --- i.e.\ our estimate for the true FRB population spectrum --- gives $\alpha=-2.29\pm0.29$, where the error represent the standard deviation of fitted residuals using a power-law shape.
Ideally, the role of weights is to perfectly re-weight \det\ to reproduce \mea\ according to Eq. \ref{recmea}.
Using Eq.~\ref{wspectrum} and \ref{unwspectrum} combined with \ref{nor1}, the weighted \det\ and \mea\ are compared in the left panel of Figure \ref{f-v-2}.
The two spectra are similar but also exhibit some differences, which will be discussed in Section \ref{5}.

\begin{figure}[ht]
\plotone{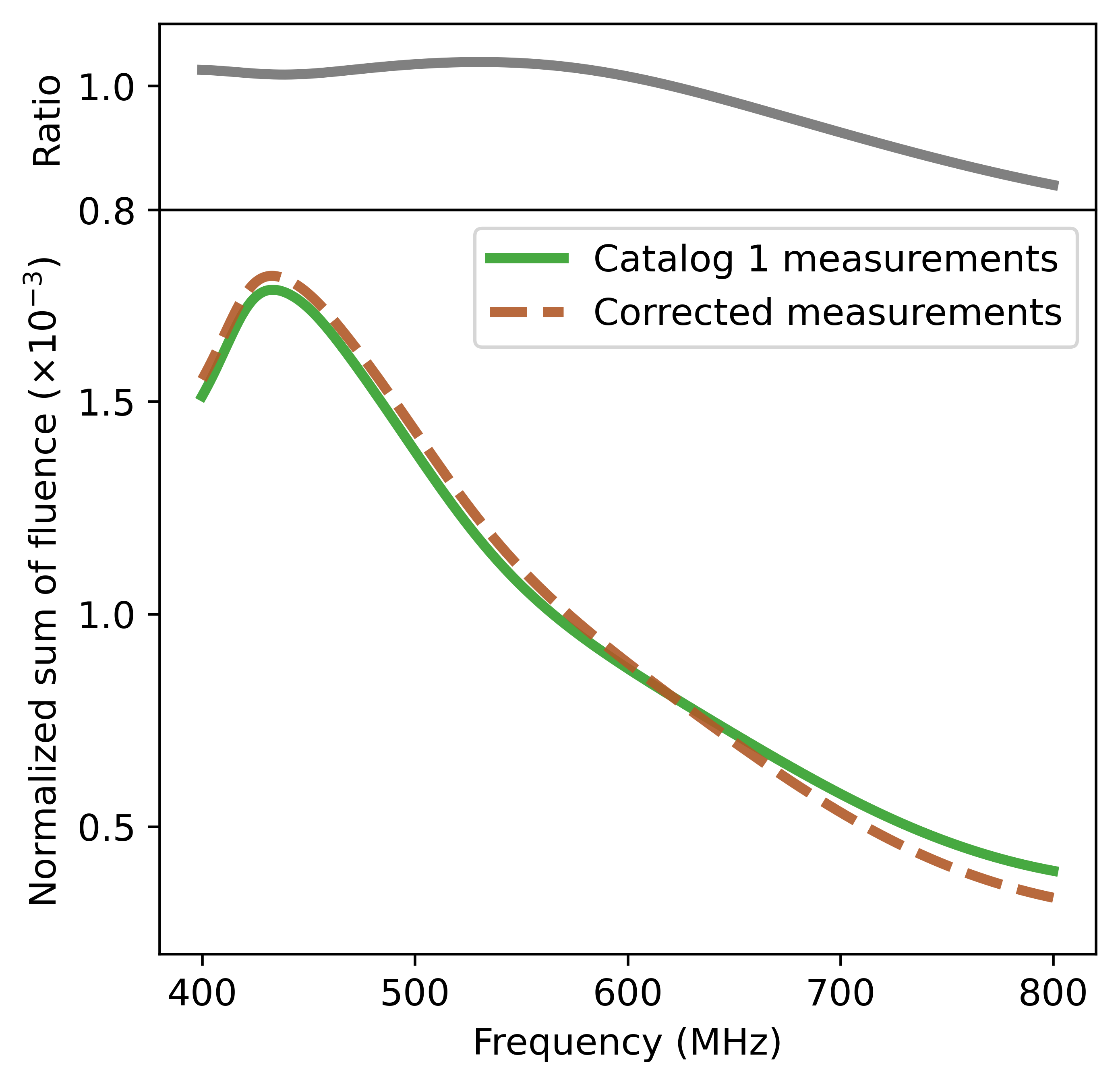}
\caption{ 
The population spectra of \mea\ and its correction.
Top panel: the gray solid curve shows ratio between weighted \tot\ and weighted \det\ in the right panel of Figure \ref{f-v-2}.
Bottom panel: the green solid curve means the spectrum of \mea. 
The brown dashed curve represents the corrected spectrum of \mea\ using the  ratio in the top panel.}
\label{f-v-meacor}
\end{figure}

We therefore also estimate the intrinsic FRB population spectrum by first calculating the frequency-dependent bias in the CHIME system from  the ratio between weighted \tot\ and weighted \det, shown in the top panel in Figure~\ref{f-v-meacor}. We then use this bias to correct the spectrum of \mea, as shown in the lower panel of Figure~\ref{f-v-meacor}.
The fitted power-law index of corrected \mea\ is $\alpha = -2.22\pm0.37$.

\begin{figure}[ht]
\plotone{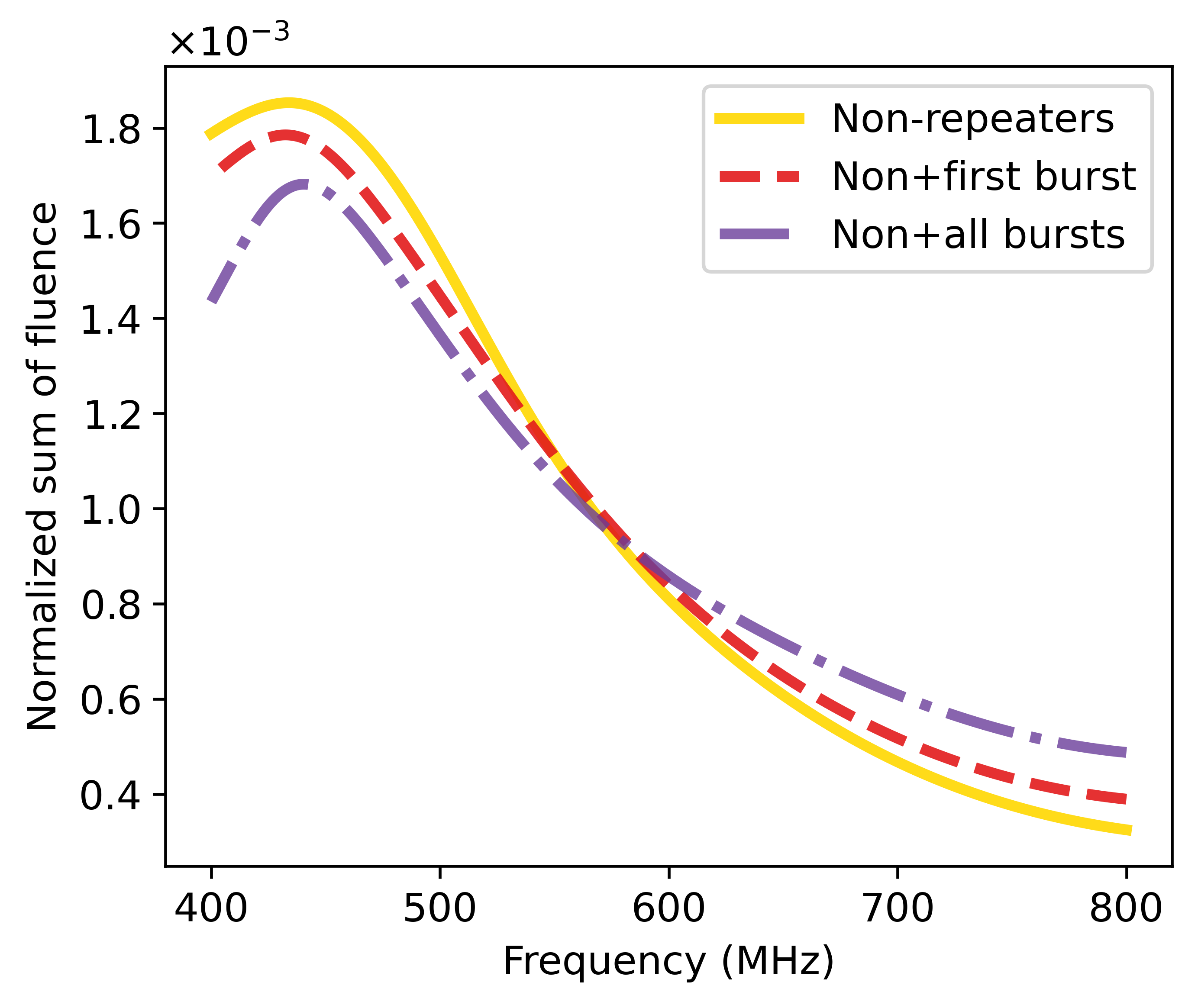}
\caption{ 
The population spectra of weighted \tot\ in different repetition cases.
The yellow solid, red dashed, and purple dashed-dotted curves represent the population spectra applying different weights calculated by: non-repeaters, non-repeaters combined with the first burst of repeaters, and non-repeaters combined with all bursts of repeaters.}
\label{f-v-renon}
\end{figure}

\subsection{Repeating and non-repeating FRBs}\label{4.3}

In the former analysis, we used non-repeaters combined with the first burst of repeaters to calculate weights as a standard procedure \citep{CHIME20a}.
Here, we investigate effects of repetition by analyzing two additional cases: only non-repeaters, and non-repeaters together with all bursts of repeaters.
The resulting population spectra from these samples are plotted in Figure \ref{f-v-renon}.
The spectrum using all bursts from repeaters ($\alpha=-1.91\pm0.20$) is flatter, and the spectrum using non-repeaters ($\alpha=-2.50\pm0.43$) is steeper, than the standard result ($\alpha=-2.29\pm0.29$).
While the former result may be due to modifications to the detection algorithm for subsequent bursts from repeaters \citep{CHIME20a}, this would not affect the first burst from repeaters. Therefore, this suggests a genuinely different frequency spectrum for repeating and apparently non-repeating bursts.

We note that differences in morphological behaviour between repeaters and non-repeaters in the CHIME sample have also been observed \citep{Pleunis21b,Sand24, Curtin24}; 
however, the properties of the high-energy part of bursts from FRB 20121102A show no significant distinction from the non-repeaters' sample \citep{Li21}.
Therefore, such differences in the FRB population may be due to a single population whereby apparent non-repeaters may be the least active repeaters, or a later evolutionary stage of them \citep{Kirsten24}; or they may be distinct populations with different progenitors and/or radiation mechanisms.

Among repeaters, two FRB sources (FRB 20180814A and FRB 20180916B) contribute to approximately half of all repeating bursts in this sample.
Therefore, the difference between results using the first burst and all bursts of repeaters mostly reflects the behaviour of very active repeaters, supporting the existence of sub-populations within repeaters, or the particular behaviour of these two sources.
Furthermore, the population spectra from non-repeaters to low-rate repeaters and then to active repeaters show a flattening trend, which may hint at  evolution between them.
However, due to differences in the detection algorithms used for the first bursts and subsequent bursts of repeaters, we cannot ensure that such differences all come from FRBs themselves.

\subsection{Narrow-band and broad-band events}\label{4.4}

\begin{figure}[ht]
\plotone{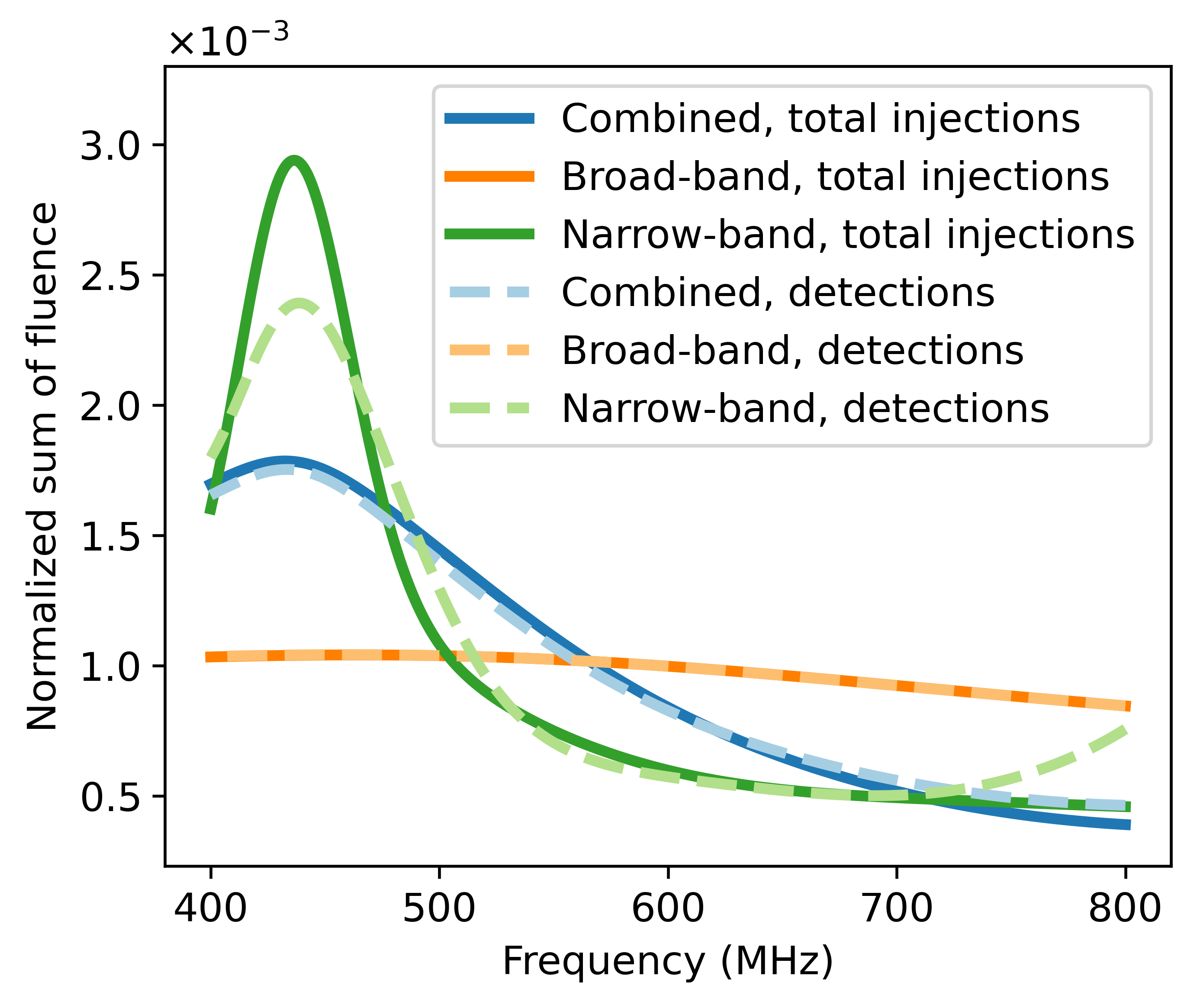}
\caption{Weighted population spectra for the cases of broad-band, narrow-band, and their combination.
The solid curves represent weighted \tot, and the dashed light curves mean weighted \det.
The orange, green, and blue curves represent the cases of broad-band, narrow-band, and their combination, respectively.}
\label{narrow-broad}
\end{figure}

In the previous analysis, we do not distinguish between narrow-band and broad-band cases. 
Here, to enhance the completeness of our analysis, we separately re-calculate the weight functions (Eq. \ref{wfinal}) for narrow-band and broad-band events and apply them to correct population spectrum, respectively.
For CHIME/FRBs, narrow-band events are defined as those with a full width at tenth maximum (FWTM) of less than 300 MHz \citep{Pleunis21a}.
Here, we simply use Eq. \ref{fgr} and true values of $(\gamma,\,r)$ to calculate the bandwidth and classify the events. 
It should be noted that using 300 MHz as a classification is a rough approximation.
Then, based on this criterion, we group \tot, \det, and \mea\ into two subsets, respectively, as well as re-organizing \rec.
For the injection data sets, roughly 70\% are narrow-band events, and for \mea, over half are narrow-band events \citep{Pleunis21a}.
Applying these sets, weighted population spectra are plotted in Figure \ref{narrow-broad}.

The results show that the narrow-band case is much steeper compared to the broad-band case, and the combined case lies between the two but closer to the narrow-band case.
The reason for the difference between narrow-band and broad-band may be due the frequency-dependent event rate.
Additionally, since narrow-band events dominates the full sample, the combined spectrum shows more toward the narrow-band case.
Meanwhile, we also notice that for narrow-band case, weighted \tot\ ($\alpha=-3.43\pm0.67$) is stepper than weighted \det\ ($\alpha=-3.02\pm0.50$), which is similar to the combined case shown in the right panel of Figure \ref{f-v-2}.
However, for broad-band case, the fitted spectrum of weighted \tot\ ($\alpha=-0.29\pm0.030$) is very close to weighted \det    ($\alpha=-0.30\pm0.029$).
This suggests that for narrow-band cases, a low-frequency bias remains exist, indicating the possibility of more events at low frequencies. 
In contrast, for broad-band cases, the detection bias is significantly smaller, and the overall spectrum still has a negative index, indicating that broad-band FRBs are brighter at low frequencies.

\section{Discussion} \label{5}

\subsection{``Down-turn" and ``up-turn" at frequency edges}\label{5.1}

In the original spectrum of \tot, we find a noticeable up-turn at the high-frequency edge. This is re-weighted by our procedure to provide a reasonable match to the data, which has power-law-like behaviour. 
However, in \tot, \det, and \mea, a down-turn at low-frequencies is found.
If this phenomenon is due to intrinsic physical characteristics, it would require a complex radiation mechanism configuration. 
Furthermore, CHIME is expected to detect FRBs up to a redshift of at least 1.5 \cite{Shin23}, meaning that any spectral features will be smeared over a similar factor in frequencies.
Therefore, we tend to believe this down-turn is non-physical and is likely due to band-edge effects \citep{CHIME22}.
Firstly, the instrument's frequency response may change rapidly at the frequency edges.
Secondly, considering that some FRBs are narrow-band, an event detected at the edge of the frequency band will have part of the burst lying outside the frequency range, reducing its detectability. This explains the downward trend at the low-frequency edge, though no such effect is seen at the high-frequency edge.

In theory, pulse injection should be able to fully account for such bias effects. 
However, in this study, the distribution of spectral properties ($\gamma$ and $r$) of \tot\ are estimated from \mea\ kernel density distribution \citep{Merryfield23}.
Due to the weighted \tot\ not providing complete freedom to reproduce an arbitrary spectral shape, the down-turn at low frequencies may be inherited from \mea, just as we see that all the measured and weighted spectra exhibit a down-turn in the low-frequency range.

To further investigate the down-turn issue, we check peak frequencies of the \tot\ and found that the peak of their distribution locates around 450 MHz.
Additionally, the number of events significantly decreases when the frequency is below 400 MHz (out-of-band events).
This means that, although the parameters of $(\gamma,\,r)$ span a wide parameter space, the resulting spectral peak frequencies are limited, indicating that injections with peak frequencies outside the band may be insufficient in the simulations.
However, the fraction of detected events which peak below the CHIME band is approximately the same as the fraction which peak in-band. This is why our re-weighting algorithm does not up-weight the spectrum at the band edge.
A explanation for why our re-weighting procedure does not adequately account for band-edge effects is that the definition of the fluence in the CHIME/FRB catalog is the in-band fluence. This definition is likely chosen due to difficulties in defining an out-of-band fluence, since spectra of the form given in Equation 1 formally integrate to infinite fluence over all frequencies. Shifting a narrow-band event partially out of band while keeping the in-band fluence constant therefore requires increasing the fluence of the event when integrated over all frequencies. Ideally, this increase in fluence would result in the event being down-weighted according the fluence weight, $W(F)$; and it would also reveal a large number of lower fluence events which would be undetectable. This would in turn result in such events being up-weighted by our procedure due to the lower detection fraction, thus correcting the band-edge down-turn. However, we cannot account for this effects here: as previously described, we do not account for correlations between $F$ and $(\gamma,\,r)$, and, furthermore, accurately defining a fluence that is not dependent on the CHIME band is difficult. We defer such an analysis to a future work.


\subsection{Frequency dependence and event rate}\label{5.2}
Our results on the frequency spectrum of CHIME/FRBs show that the \mea\ have more power or higher event rate at low frequencies, and that the true population is even more low-frequency biased. 
\cite{James22} discussed extreme interpretations of frequency spectrum: the “spectral index” interpretation, where all FRBs are broad-band and have more power at low frequencies; the “rate” interpretation, where all FRBs are narrow-band and there are more events at low frequencies. 
The spectral description developed by \cite{CHIME21a} (Eq. \ref{fgr}) lies somewhere in the middle of these interpretations.
Therefore, the obtained population spectral index in this work can be linked to both the total power and the event rate index.

Our intrinsic spectral index of $\alpha = -2.29\pm0.29$ is significantly steeper than that derived by other studies. 
Using the spectral index interpretation and 23 measured values from ASKAP/FRBs, \cite{Macquart19} obtained a population spectral index of $\alpha = -1.5^{+0.2}_{-0.3}$.
Applying the rate interpretation, \cite{James22} corrected \cite{Macquart19}'s index to $\alpha = -0.65 \pm 0.3$.
The differences may be caused by different frequency bands of ASKAP and CHIME, resulting in distinct characteristics. 
Meanwhile, differences in instrument characteristics like flux threshold, beam shapes, selection effects can also lead to varying observational biases.

In \cite{Bhattacharyya22}'s work, they modeled FRB population and obtained the best fitting index is $\alpha = -1.53^{+0.29}_{-0.19}$. 
\cite{Shin23}'s work analysed \mea\ to model the energy and distance distributions.
The derived index is $\alpha = -1.39^{+0.86}_{-1.19}$, which is flatter than ours, but consistent at the $1\,\sigma$ level.
The reasons for the differences between ours and theirs can be due to the model dependence in FRB population modeling, whereas we used real measured and injection data.
For example, population modeling is related to assumptions of source evolution because redshift evolution and the intrinsic evolution of FRBs are degenerate.
Meanwhile, in those modeling studies, all FRB events are assumed to share a flat power-law spectrum, which is another form of model dependence.
In contrast, using real data avoids these dependency issues, as the spectrum of each event is determined by distinct spectral parameters. 
For instance, some events can span the entire frequency range, while some events are close to a delta function form.
It is worth noting that although our analysis is less model-dependent, it is not entirely model-free. 
We construct the true fluence distribution based on a flat local universe, and our resulting spectrum (which we fit with a power-law) has unphysical down-turns near the edges, as discussed in Section \ref{5.1}.
This treatment is however consistent with FRB population modelling \citep{CHIME21a, James22}, so our results should still be useful input for those works.




Finally, a question arises: is this consistent with low-frequency FRB observations and limits?
The lowest observed frequency of FRBs is at 110 MHz for FRB 20180916B from Low Frequency Array (LOFAR) \citep{Pleunis21b}.
\cite{Pastor21} showed that FRB 20180916B is more active below 200 MHz than at 1.4 GHz.
However, CHIME has shown that their event rate \citep{CHIME21a} is approximately the same as ASKAP \citep{Macquart19}.
As discussed in the introduction, difficulties arise from various aspects \citep{Pilia21}, such as scattering and dispersion smearing, RFI challenges, and signal absorption from the source's environment.
These challenges explain why, despite evidence suggesting that FRBs are stronger or have higher rates at low frequencies, the actual detection rate of low-frequency remains low.

\section{Conclusion} \label{6}

In this work, applying \mea\ and its injection system, we expand the correction method of CHIME/FRB to build a statistical model. 
After accounting for selection effects and correlations, we further correct the CHIME/FRB population distribution and, for the first time, obtain the population spectral characteristics for the entire Catalog 1 sample. 
The main conclusions of this work are summarized:
\begin{itemize}
    \item Based on the bias-corrected method, population spectrum of weighted \tot\ is obtained with a power-law index of $\alpha=-2.29\pm0.29$, which can be seen as the intrinsic population spectrum within CHIME frequency band.
    \item The ratio between weighted \tot\ and weighted \det\ represents a frequency-dependent correction, revealing a bias at low frequencies for CHIME/FRBs.  Applying the correction ratio to the population spectrum of \mea, a steeper spectrum is obtained, suggesting that FRBs are brighter or have higher event rate in the low-frequency region.
    \item Using the same bias correction method, we investigate the effects of FRB repetition.
    The corrected population spectra for non-repeaters, non-repeaters combined with the first bursts of repeaters, and with all bursts of repeaters show different fitted indices. 
    This result supports the hypothesis that non-repeaters, low-rate repeaters, and active repeaters have different progenitors, radiation mechanisms, or evolutionary stages.
    \item Applying the final multi-weight function, distributions of DM, pulse width, and scattering timescale are updated.
    The corrected distributions of DM and pulse width are generally consistent with CHIME's corrections.
    However, in the correction of scattering timescale, our result suggests that there are more short-scattering events, which contrasts with CHIME's correction that indicates more long-scattering events should exist.
    The differences may be due to our inclusion of two spectral properties $(\gamma,\,r)$ and parameter correlations in the analysis, as well as correlations between $(\gamma,\,r)$ and scattering.
\end{itemize}

\begin{acknowledgments}
We thank Kiyoshi W.\ Masui and Kaitlyn Shin for discussions of CHIME's bias-correction method, Ziggy Pleunis for the suggestion of separating the sample into frequency broad-band and narrow-band components, discussions with the CIRA/FRB team and the NAOC/ISM team, and the referee for their valuable suggestions.
This work is supported by the National Natural Science Foundation of China No. 11988101, the National Key R\&D Program of China No. th, the International Partnership Program of Chinese Academy of Sciences No.114A11KYSB20210010, grant NSF PHY-2309135 to the Kavli Institute for Theoretical Physics, and Di Li is a New Cornerstone investigator.
XHC acknowledges the support from China Scholarship Council (Grant No. 202304910441).
CWJ acknowledges support from the Australian Government through the Australian Research Council Discovery Project DP210102103.

\end{acknowledgments}


\appendix
\section{Corrected spectra with data filtering }\label{A}
Data filtering is commonly used in previous studies, with the most mentioned reason being to reduce potential errors in Catalog 1 measurements \citep{CHIME21a, Hashimoto22, Shin23}. 
In our analysis, we did not apply data cuts in the main analysis \citep{Chen22}, as we wanted to keep more data and avoid introducing additional human selections into our model. 
To check the impact of this decision, we applied the filtering method to re-calculate the weight function and population spectra using the same procedure in Section \ref{3}.
Here are the criteria we used for making cuts both on \mea\ and injection data sets, and events meeting these criteria will be removed: 
\begin{enumerate}
    \item Low-S/N  events with $\mathrm {S/N<12}$.
    \item Events with $\mathrm {DM<1.5\times max(DM_{NE2001},\, DM_{YMW16}})$ or $\mathrm{DM<100\,pm\,cm^{-3}}$.
    \item Events with scattering timescale $\mathrm {\tau >10\,ms}$ at 600 MHz.
    \item Events detected in far side-lobes. For injections, we applied the coordinate $\mathrm{\left| x \right|>1.5}$ of CHIME/FRB beam model to act as the outside the primary beam \citep{CHIME20a}.
\end{enumerate}
After applying these filters, 290 Catalog 1 events (279 non-repeaters and 11 first burst of repeaters), $\sim10^6$ total simulated injections, and $\sim 1.6\times 10^4$ detected injections are survived.
\begin{figure}[ht]
\plotone{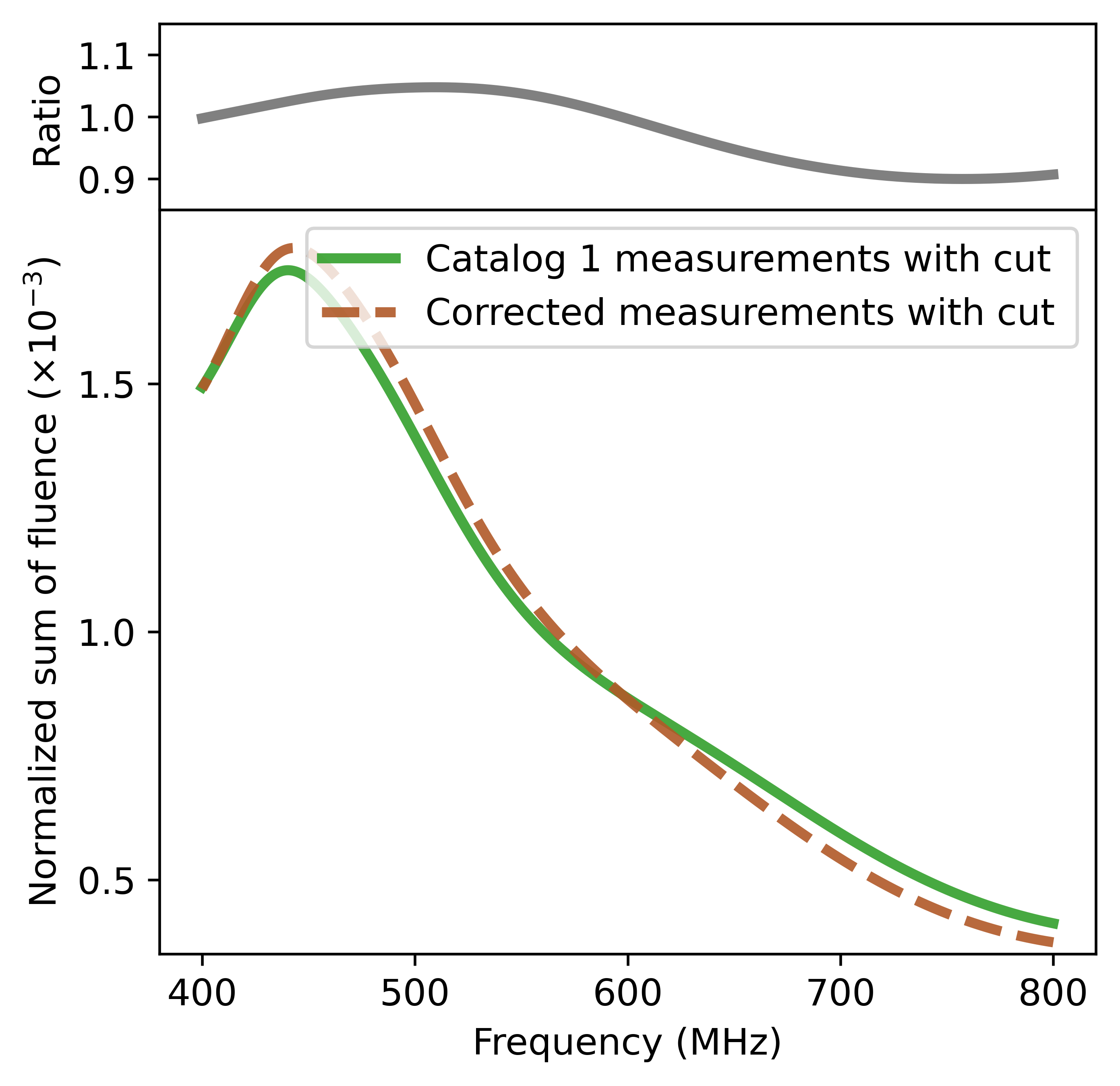}
\caption{ 
With data filtering applied, the population spectrum of \mea\ and its correction. 
Other notes are the same as in Figure \ref{f-v-meacor}, except that here the corrections are based on the cut data.}
\label{f-v-meacor-cut}
\end{figure}

We re-calculated the final multi-weight function Eq. \ref{wfinal} using the cut data sets and applied it to the \tot\ spectra, following the same approach as described in the main text.
Subsequently, we use the weighted \tot\ and weighted \det\ to calculate the ratio as shown in the upper panel in Figure \ref{f-v-meacor-cut}.
Then, we apply this ratio to the \mea\ to obtain the corrected spectrum ($\alpha=-2.15\pm0.36$), as shown in the lower panel of Figure \ref{f-v-meacor-cut}.
The correction trend is consistent with that filtering is not considered (Figure \ref{f-v-meacor}), which also suggested that there should be more events or FRBs are brighter at the low frequencies. 
Meanwhile, this also indicates that using different injection data sets does not significantly change our conclusions.

\section{Comparison with CHIME corrected distributions in 4-dimensional case}\label{B}

In Section \ref{4.1} of the main text, we presented the distributions of dispersion measure (DM), pulse width, and scattering timescale with considering the full final multi-weight of $W(\mathrm{DM}, w, \tau, \gamma, r, F)$, finding that our corrected distributions of scattering timescale show significant differences compared to CHIME's corrections. 
We speculate that these differences are primarily due to the inclusion of two spectral parameters ($\gamma,\,r$) in the weight calculation.
Therefore, we conducted further analysis in this appendix.

\begin{figure}
\plotone{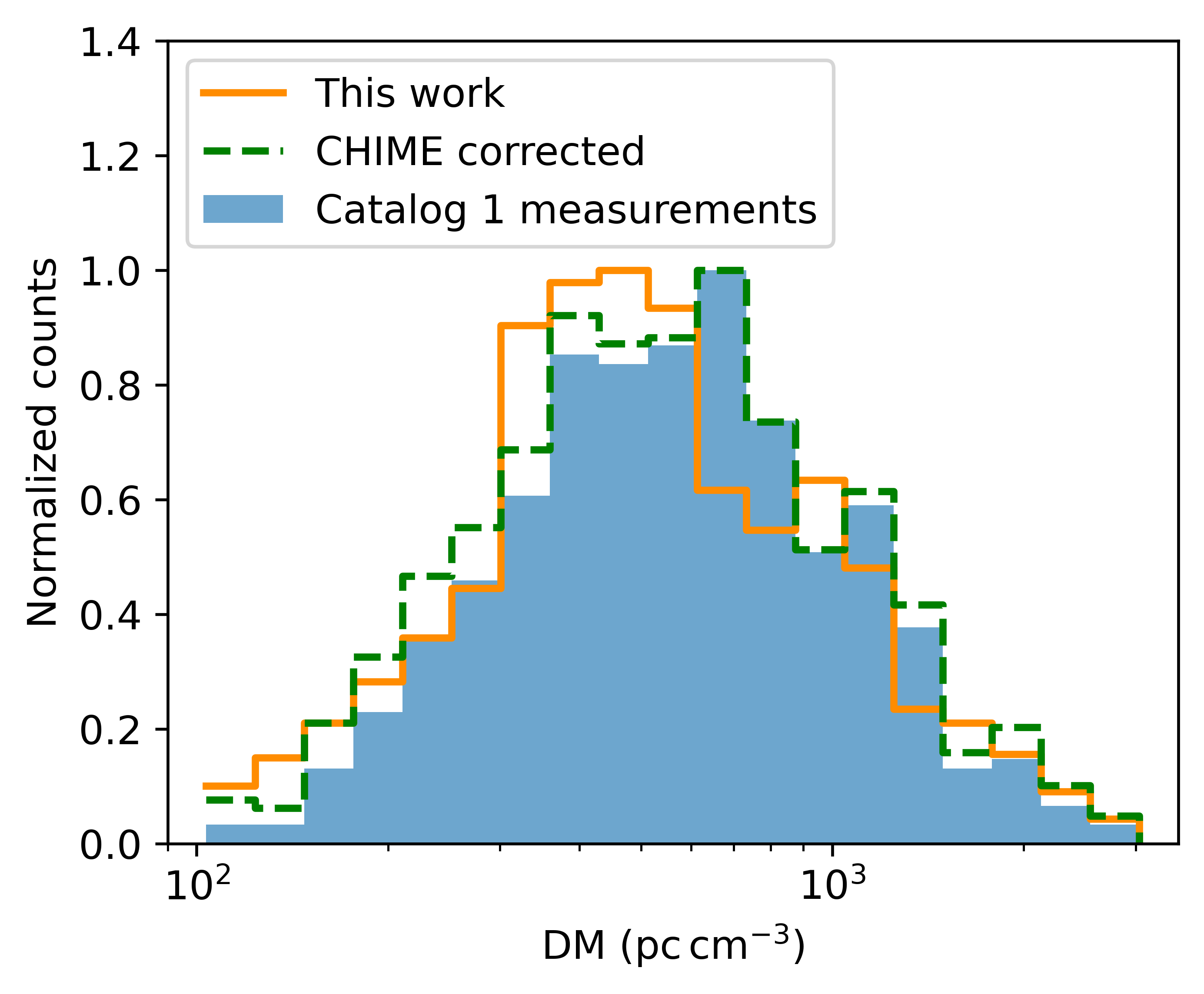}
\caption{Distribution of dispersion measure (DM) with considering different correction method.
The orange solid distribution represents the corrected \tot\ using the 4-dimensional weight function in Appendix \ref{B}.
Here, we use the maximum value for normalization.
Other annotations are consistent with those in Figure \ref{dm-dis}.}
\label{dm-dis-4d}
\end{figure}

\begin{figure}
\plotone{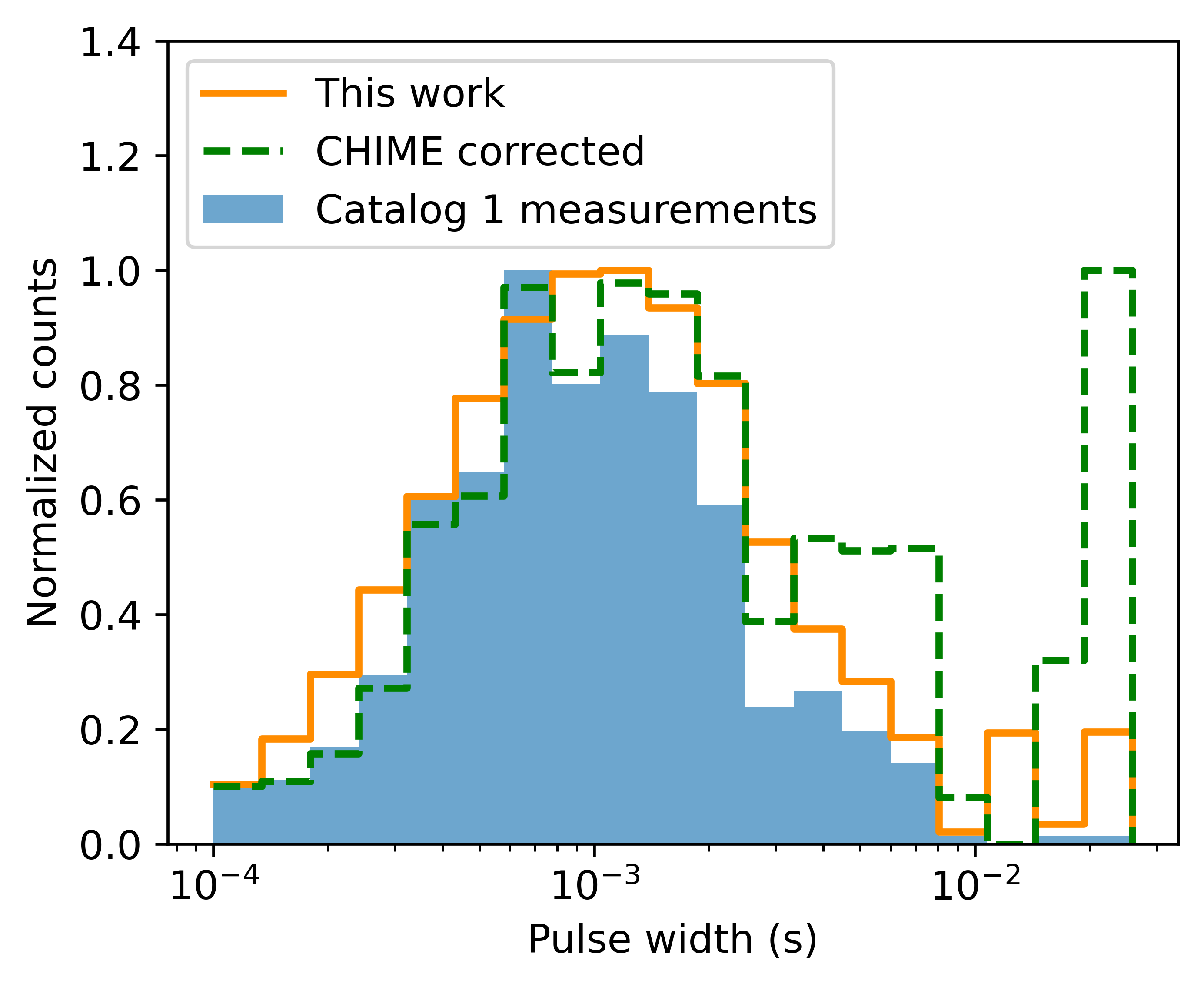}
\caption{Distribution of pulse width ($w$) with considering different correction method.  
The orange solid distribution is obtained from the 4-dimensional weight.
Other annotations are consistent with those in Figure \ref{w-dis}.}
\label{w-dis-4d}
\end{figure}

\begin{figure}
\plotone{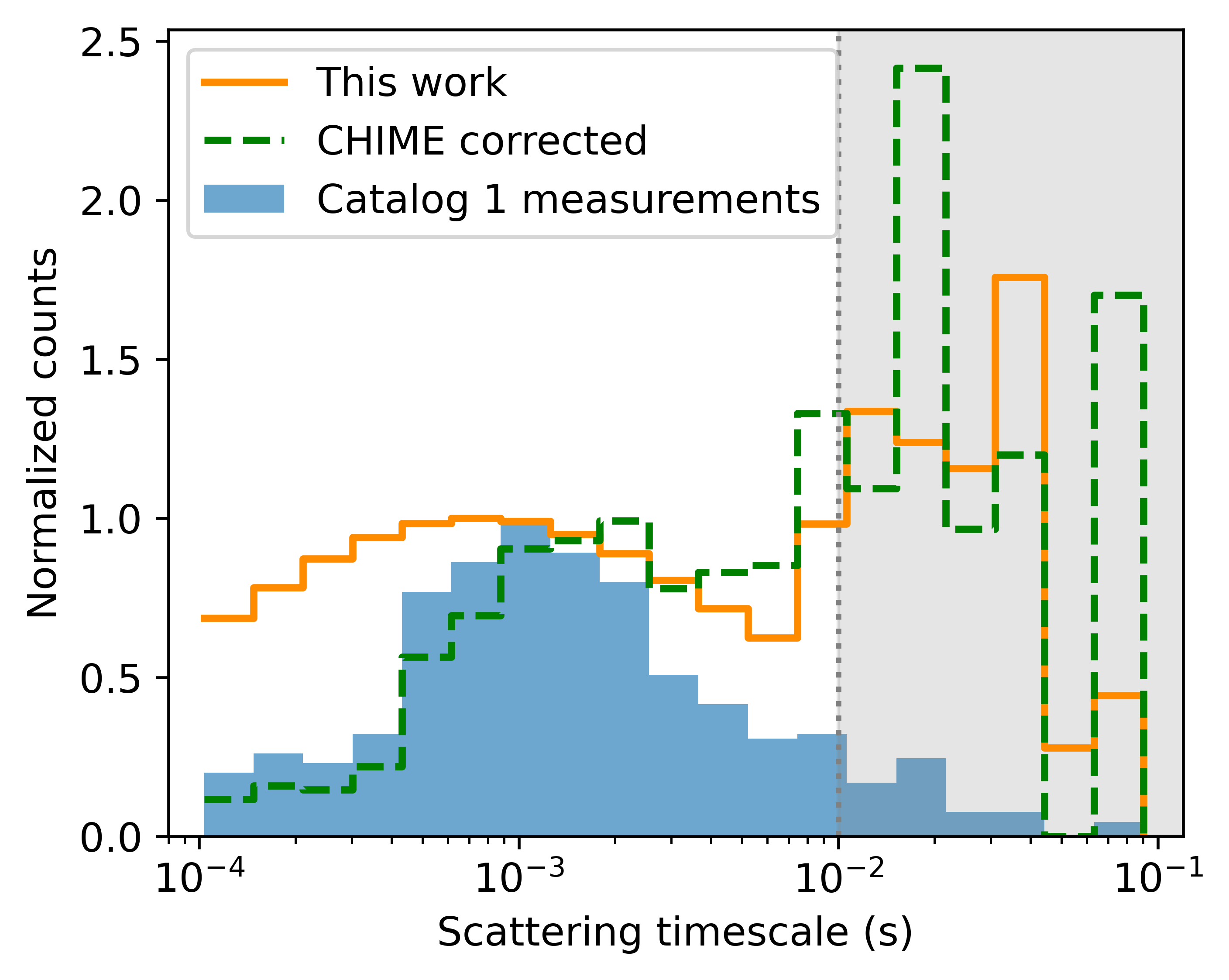}
\caption{Distribution of scattering timescale ($\tau$) with considering different correction method. 
The orange solid distribution is obtained from the 4-dimensional weight.
Other annotations are consistent with those in Figure \ref{tau-dis}.}
\label{tau-dis-4d}
\end{figure}

In the work of \cite{CHIME21a}, they independently calculate selection functions considering DM, pulse width, scattering timescale, and fluence, and then multiplied them together for the correction.
Therefore, we also calculate the weights considering only these four parameters. 
However, to ensure model consistency described in Section \ref{3}, we still accounted for the correlations, that is calculating the 4-dimensional weight as
\begin{equation}
    W(\mathrm{DM}, w, \tau, F)=W(\mathrm{DM}, w, \tau)\times W(F).
    \label{weight4}
\end{equation}
Subsequently, after iterations the corresponding final weight is applied to the \tot\ to obtain the corrected distribution as plotted in Figure \ref{dm-dis-4d}, \ref{w-dis-4d}, and \ref{tau-dis-4d}.
The results show that when only considering these four parameters for corrections, our results are very close to those obtained from CHIME's corrections.
Although some minor differences remain, these are likely due to statistical fluctuations or  consideration of correlations.
Thus, this supports the explanation that the differences presented in Section \ref{4.1} are primarily due to the inclusion of spectral parameters in the full multi-weight $W(\mathrm{DM}, w, \tau, \gamma, r, F)$.

\begin{figure}[ht]
\plotone{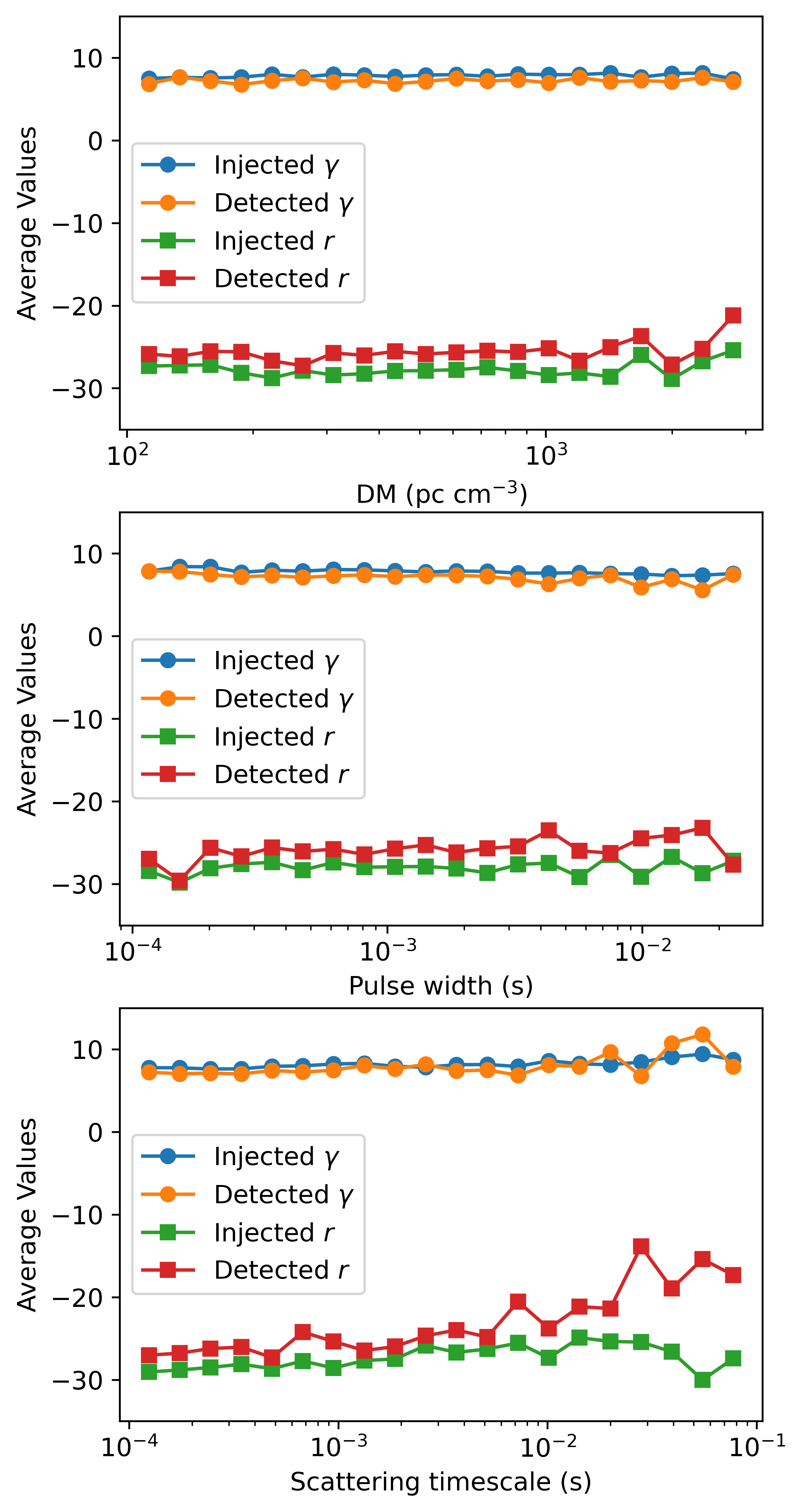}
\caption{Mean value variation plots of spectral index $(\gamma)$ and spectral running $(r)$ in the parameter spaces of dispersion measure (DM) (upper panel), pulse width (mid panel), and scattering timescale (bottom panel) for both \inj\ (injected) and \det\ (detected).
The blue and orange lines represent injected and detected mean values of $\gamma$, respectively.
The red and green lines mean injected and detected average values of $r$, respectively.
The circular and square markers represent the bin centers for each parameter, consistent with the histograms shown in Figures \ref{dm-dis-4d}, \ref{w-dis-4d}, and \ref{tau-dis-4d}.}
\label{gr-3dspace}
\end{figure}

\begin{figure}[ht]
\plotone{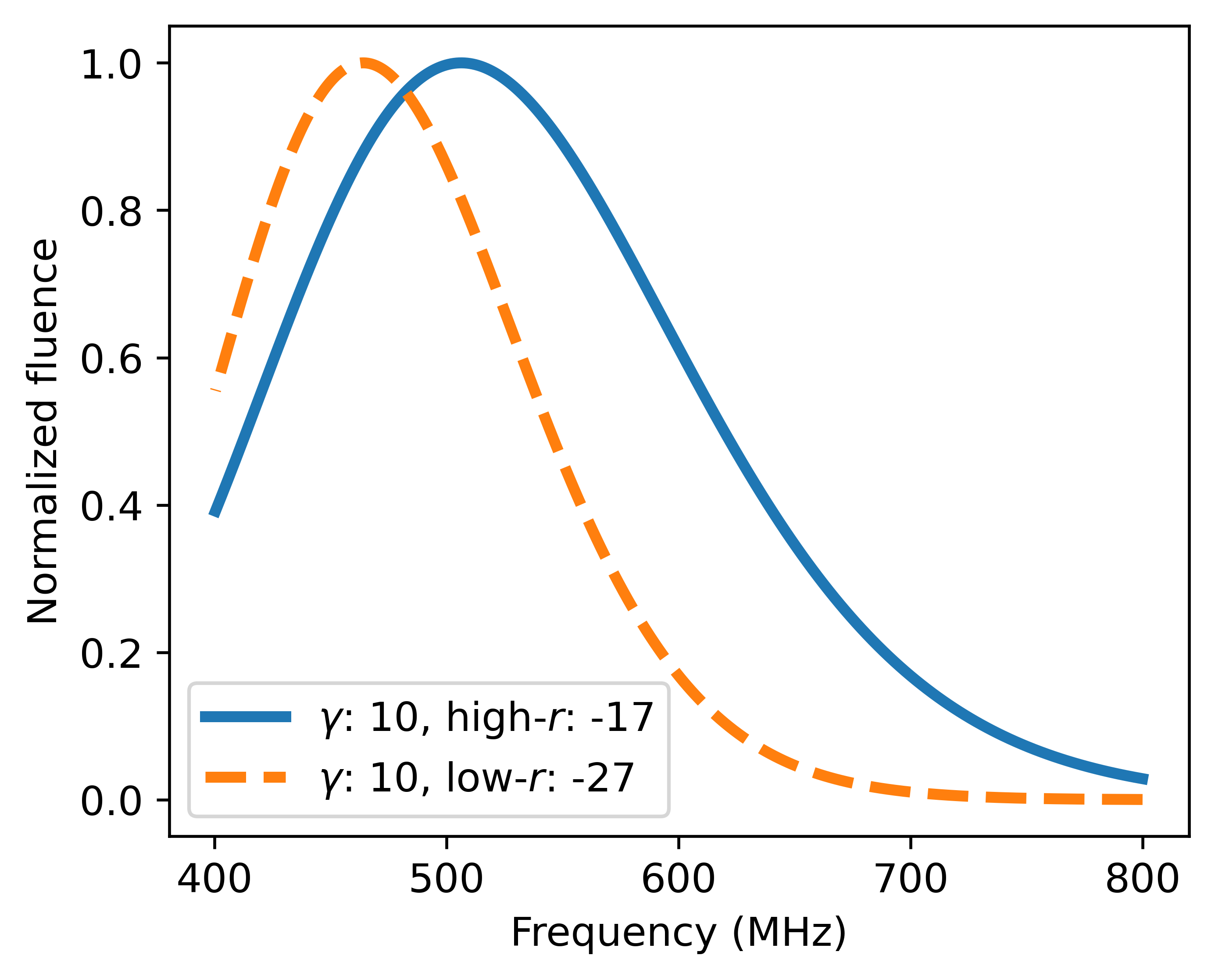}
\caption{Schematic diagram of events' spectra with high-$r$ (blue solid curve) and low-$r$ (orange dished curve) based on Eq. \ref{fgr}.
We keep $\gamma$ consistent for better comparison. 
The specific values of high-$r=-17$ and low-$r=-27$ are taken from the average $r$ values in the longest scattering timescale bin in the bottom panel of Figure \ref{gr-3dspace}.}
\label{lowhigh_r}
\end{figure}

To further investigate why $(\gamma,\,r)$ significantly affects the corrected distributions of the parameters, we plotted the variation in the average values of $(\gamma,\,r)$ in the parameter spaces of DM, pulse width, and scattering timescale for both \inj\ and \det\ as shown in Figure \ref{gr-3dspace}.
We notice that in the scattering timescale space, the average values of $r$ show significant deviations in the long-scattering regions between \inj\ and \det, while the differences are relatively small in other cases.
This may indicate that the CHIME/FRB detection pipeline has a bias toward high-$r$ events in the long-scattering space.
The high-$r$ event represents broader frequency bandwidth and higher peak frequency, as illustrated in Figure \ref{lowhigh_r}.
The exact reasons for this kind of bias in the detection system remain unclear to us and are not the focus of this work, but this issue is worth to be noted.

\bibliography{references}{}
\bibliographystyle{aasjournal}

\end{CJK*}
\end{document}